Facultad de Biología

Universidad de La Habana

# LA FAUNA DE MAMÍFEROS FÓSILES DEL DEPÓSITO PALEONTOLÓGICO “ EL ABRÓN" (NIVEL IX), PINAR DEL RÍO, CUBA.

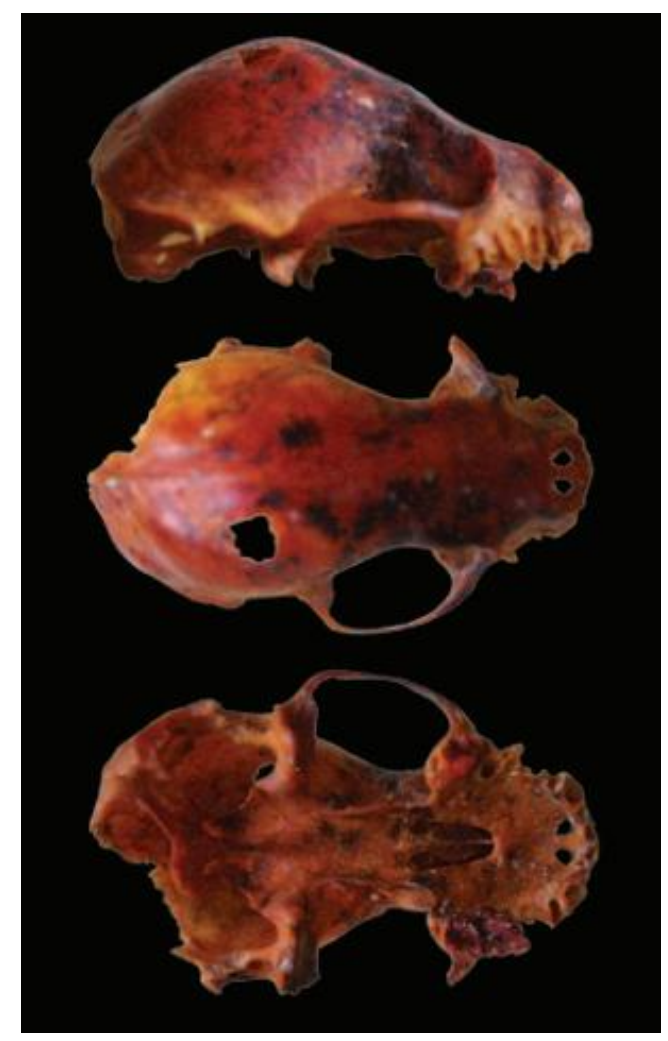

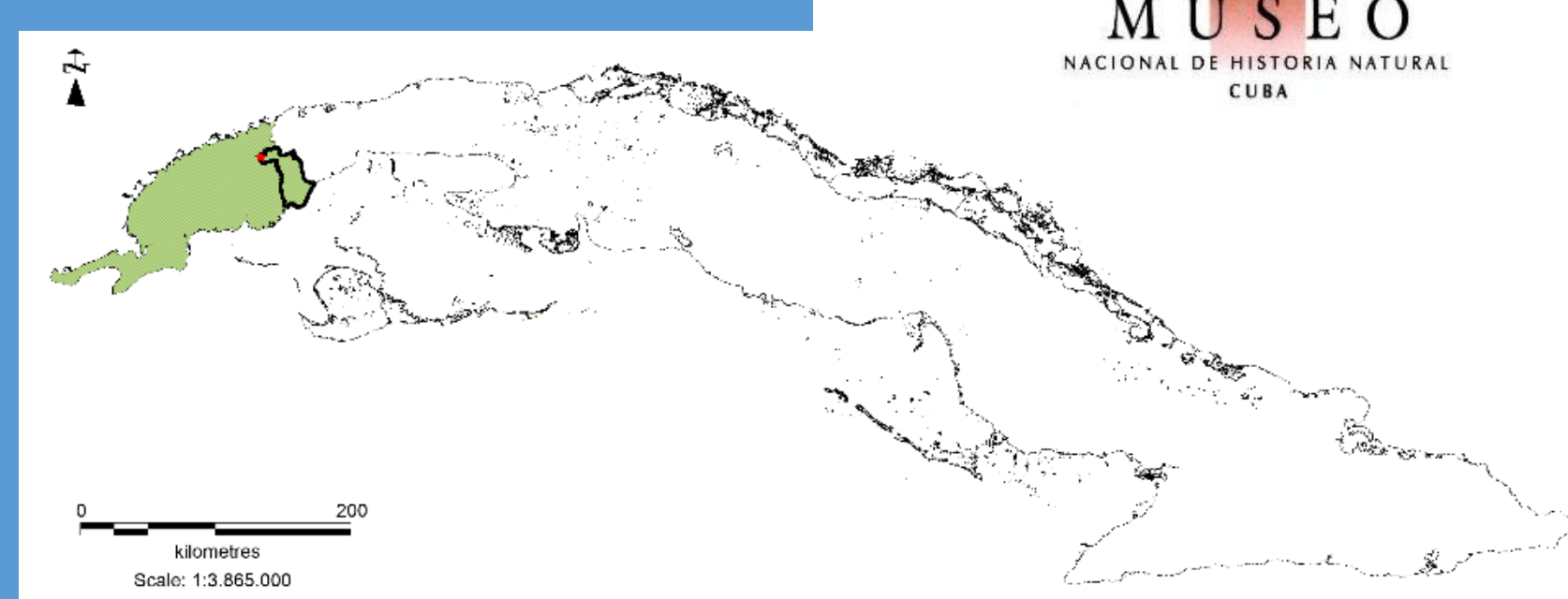

Autora:

Soraida Fiol González

Tutor:

MSc. Joao G. Martínez L.

**UNIVERSIDAD DE LA HABANA**

**FACULTAD DE BIOLOGÍA**

**Carrera de Biología**

**TÍTULO**:

# LA FAUNA DE MAMÍFEROS FÓSILES DEL DEPÓSITO PALEONTOLÓGICO “ EL ABRÓN" (NIVEL IX), PINAR DEL RÍO, CUBA.

**Autora**:

Soraida Fiol González

**Tutor**:

MSc. Joao G. Martínez López.

Museo Nacional de Historia Natural, La Habana, Cuba (MNHNCu). Obispo no. 61, Plaza de Armas, La Habana 10100 CUBA.
Correo: jgml@mnhnc.inf.cu

Esta tesis se defiende en la Facultad de Biología de la Universidad de La Habana.

*A mami, papi y tata*

**“La Paleontología y la Geología aún le guardan muchos secretos a nuestros investigadores, y parece no estar lejano el día en que determinados descubrimientos, estremezcan los cimientos de los dogmas establecidos y hagan variar sus ideas, que parecen eternamente fijas, como si fueran movidas bruscamente por un poderoso resorte”**

**Oscar Arredondo (1958:12)**

## *AGRADECIMIENTOS*

Mi más eterno agradecimiento al MSc. Joao G. Martínez López, por dedicar parte de su preciado tiempo a la tutoría de esta tesis, gracias por su enorme paciencia, por confiar en mí, por todo el apoyo brindado y por los buenos consejos siempre que los necesité.

A Ernesto Aranda Pedroso, por haber aceptado la revisión crítica de este documento mediante su oponencia.

A todos los investigadores que nos ayudaron con la identificación de los restos óseos: Dr. Carlos Arredondo Antúnez de la Facultad de Biología, Gilberto Silva Taboada del Museo Nacional de Historia Natural de Cuba, Osvaldo Jiménez del Gabinete de Arqueología y Lazarito William.

Al Dr. Reinaldo Rojas, al Dr. Rafael Borroto Páez y a Luis Martínez Zamora (Historiador del Municipio Los Palacios, Pinar del Río), por facilitarme la información y la confianza expresada en la fructificación de este trabajo.

A Emilio, por dedicar alguna que otra hora a ayudarme con la selección y clasificación del material.

A Eduardo, por sacrificar tantos sábados para asistir a la Biblioteca Villena a auxiliarme en la confección de esta tesis.

A Mei por toda la atención y hospitalidad prestada, sobre todo en estos últimos días.

A mis amigos más cercanos, Lizzette, Ariadna, Claudia, Aimee, Heydi, Camilo, Johana y Víctor, por haber estado juntos toda la carrera, enfrentando las cosas malas y celebrando las buenas, pero siempre dispuestos de ayudarnos el uno al otro sin importar las dificultades.

A mi buena amiga Taisia, por estar siempre al tanto de cada paso que daba y por alentarme a seguir adelante en todo momento.

A mis profesores: Wilfredo, Martha Pérez, Alina, Gloria, Rosalina, Annele, Junier, Dennis Denis, Ana Sanz y al resto del claustro de profesores de la Facultad de Biología por toda la preparación y el conocimiento brindado durante todos estos años.

A Sara, Nilda, Elsa, Lourdes y al resto del personal de la secretaría por todo su apoyo y comprensión durante estos largos años de mi carrera.

A Gissela, Leo, Aralia, Naibi, Vidar, Merci, Silvia, Juan, Emigdio, Jesús, Juan Carlos, Charo, Fefita, Fredy, Marcel, Joel, Javi, Leito, Ewar, Tere y Haniti quienes me estimularon y alentaron a no desistir y confiar en que llegaría hasta el final brindándome todo su apoyo. A todos les muestro mi gratitud incondicional y mi cariño infinito.

A mis padres que estuvieron ahí en todo momento alentándome o peleándome (sobre todo mi mami), soportando mi mal humor y aconsejándome.

A mi tata que a pesar de estar lejos siempre estuvo ahí para mí.

A Rami, por su apoyo incondicional, paciencia, comprensión, soportarme en estos últimos meses, por sus mimos y por quererme tanto.

A todos aquellos que aportaron todo su esfuerzo e interés, en la realización de tan importante investigación quiero que sientan desde esta página mi más sentida gratitud.

**RESUMEN**

El depósito fosilífero "El Abrón", ubicado en Pinar del Río, Cuba, y de cuya antigüedad solo se tiene como referencia el nivel VII (17 406 años AP), está clasificado como el mayor conjunto de fósiles acumulados para nuestro archipiélago, producidos por la acción trófica de lechuzas durante miles de años. Abundante material óseo procedente de este depósito requiere ser estudiado en aras de identificar la fauna de vertebrados que formaba parte de la dieta de las estrígidas del pasado. El objetivo de este trabajo fue determinar la composición taxonómica de la fauna de mamíferos extinguidos y vivientes, mediante el estudio paleontológico del nivel de mayor profundidad de dicho depósito (Nivel IX). Se analizó el material extraído que se encuentra actualmente depositado en el almacén de colecciones paleontológicas del Museo Nacional de Historia Natural, La Habana, Cuba (MNHNCu). Se procedió a la limpieza de los restos óseos, clasificación e identificación de las especies, así como análisis tafonómico del estado de conservación de los restos. Se encontró que la fauna de mamíferos del depósito paleontológico objeto de estudio está compuesta esencialmente por 3 órdenes, 7 familias y 14 especies. El orden más representativo es Chiroptera (fauna de murciélagos), representada por 4 familias, 9 géneros y 9 especies del total identificadas. Se reportaron por primera vez para la localidad cuatro especies de murciélagos *Erophylla sezecorni, Monophyllus redmani, Pteronotus parnelli* y *Tadarida brasiliensis.* Estos resultados constituyen la base para futuros estudios paleoecológicos que permitan reconstruir la historia natural de estas especies. Además, el hallazgo de nuevas especies en esta zona constituye un aporte al conocimiento acerca de la distribución de estas especies en el archipiélago cubano y la antigüedad de las mismas. Finalmente, el análisis tafonómico del estado de conservación de estos restos permitió la comprensión de los procesos que dieron origen al depósito y las características del mismo, lo que contribuye a una estimación adecuada de las especies presentes en el mismo y la relación espaciotemporal de los restos.

**ABSTRACT**

"El Abrón" is a fossil deposit located in Pinar del Rio, Cuba, and whose age is only reference level VII (17 406 years BP), it is classified as the largest collection of fossils accumulated for our archipelago, produced by trophic action of barn owls for thousands of years. Abundant bone material from this deposit needs to be studied in order to identify vertebrate species that it was part of the diet of owl in the past. The aim of this study was to determine the living taxonomic composition of the fauna of extinct mammals, and throughout the paleontological study of the deeper level of said tank (Level IX). The extracted material which it is currently stored in the warehouse of paleontological collections of the National Museum of Natural History in Havana, Cuba (MNHNCu) was analyzed. We proceeded to clean the bones, to classify and to identify them from the species and also the taphonomic analysis of the condition of the remains. It was found that the mammal fauna of the paleontological deposit under study is composed essentially of 3 orders, 7 families and 14 species. The most significative order is Chiroptera (bat fauna), represented by 4 families, 9 genus and 9 species of the total which were identified. There were reported four species of bats *Erophylla sezecorni*, *Monophyllus redmani*, *Pteronotus parnelli* and *Tadarida brasiliensis* in the location. The results are the basis of the future paleoecological studies in order to reconstruct the natural history of these species. Moreover, the discovery of new species in this area is a contribution to the knowledge about the distribution of these species in the Cuban archipelago and the age of them. Finally, the taphonomic analysis of the conservation status of these remains permitted the understanding of the processes that gave rise to the tank and its characteristics, and also it contribute to an adequate estimation of the species present in it and the relationship between spatiotemporal with the fossil.

**ÍNDICE**

## INTRODUCCIÓN

La naturaleza cársica de nuestro país favorece la formación de depósitos que actúan como embudo de captación de los restos de especies, las cuales con el tiempo pasan a ser subfósiles y finalmente fósiles, si los ambientes diagenéticos con los que interactúan los conducen a este proceso. La Sierra de la Güira presenta una estructura geológica típica de carso de pequeñas montañas donde grietas, casimbas, solapas, cuevas y oquedades se desarrollan producto de la interacción de los procesos hidrogeoquímicos con el sustrato geológico, relacionados con cambios paleoambientales. Sin embargo, algunas de las estructuras o formaciones cársicas favorecen la actividad ecológica de especies, dentro de las cuales podemos mencionar a las lechuzas.

La importancia del estudio de reservorios de aves permite conocer aspectos de la biodiversidad de Cuba tanto actual como pasada, aspecto este muy importante para los estudios paleontológicos, sobre todo de fauna de microvertebrados. Se conoce que los depósitos fosilíferos cubanos de edad Cuaternario son los que fundamentalmente contienen los registros de la acción trófica de aves fósiles, y que éstos se remontan desde la época Pleistoceno. La composición de las regurgitaciones (*egagrópilas*) que han quedado en el registro estratigráfico como evidencia de su acción trófica están consideradas de gran importancia, tanto para estudios taxonómicos y ecológicos actuales, como paleontológicos.

Se conoce además que el género *Tyto* (lechuzas), es el responsable de residuarios paleontológicos importantes durante el Pleistoceno – Holoceno, ya que sus hábitos nocturnos les brindan ventajas en sus relaciones interespecíficas. La captura de sus presas y la conservación de los restos que quedaron en sus residuarios es una evidencia directa de la paleoecología del período cronológico en el que habitaron.

El depósito fosilífero "El Abrón" está clasificado como el mayor conjunto de fósiles acumulados para nuestro archipiélago, producidos por la acción trófica de lechuzas (vivientes y extinguidas) durante miles de años. Hasta el momento la única publicación que recoge los resultados científicos a partir de materiales fósiles recuperados del mismo es la de Suárez y Díaz-Franco (2003), donde se describe una nueva especie de murciélago para Cuba (*Phyllops silvai*), y se anuncia, por primera vez, la presencia de millones de restos óseos acumulados de microvertebrados, tanto vivientes como extinguidos.

Este importante yacimiento se encuentra localizado en el municipio de Los Palacios de la occidental provincia de Pinar del Río. Dentro de los paisajes montañosos más conocidos para esta región es la Sierra de los Órganos dentro de la cordillera de Guaniguanico la que incluye el área del depósito. La denominada Sierra de la Güira, es una pequeña porción montañosa sin aparente conexión con otras estructuras geológicas similares que actúa y ha actuado durante miles de años como receptor de restos de especies que habitaron y habitan el archipiélago cubano.

El depósito "El Abrón" marca un período cronológico importante para los estudios paleoecológicos y paleogeográficos en Cuba, pues un fechado radiocarbónico sobre material óseo de una lechuza extinta (*Tyto noeli*) ubica el estrato estudiado en 17 406 ±161 años AP. El material rescatado se encuentra depositado en el Museo Nacional de Historia Natural, La Habana, Cuba [MNHNCu], donde abundante material óseo requiere ser estudiado en aras de identificar la fauna de vertebrados que formaba parte de la dieta de las estrígidas del pasado. Esta acumulación de carácter natural cuenta con nueve niveles dentro de los cuales solamente el séptimo ha sido estudiado sobre la base del fechado mencionado, y de la descripción de la especie *Phyllops silvai*.

Nuestro principal propósito en este trabajo es estudiar el nivel estratigráfico de mayor profundidad, asumiendo una mayor antigüedad en el mismo para todos los taxones identificados, teniendo en cuenta su disposición estratigráfica sin alteraciones antropogénicas (Suárez y Díaz-Franco, 2011)[1], y así determinar la composición general de la fauna de micromamíferos fósiles y vivientes.

También se pretende valorar el estado de conservación del depósito para correlacionar las características de los materiales hallados y el resultado taxonómico obtenido. Se trata de conocer, en principio, la composición taxonómica del estrato más antiguo del depósito, aportando nuevos datos paleontológicos a la reconstrucción de la historia natural de las especies que habitaron y habitan el archipiélago cubano.

---

[1] SUÁREZ, W. Y DÍAZ-FRANCO, S. (2011). Estudio paleontológico del depósito fosilífero El Abrón, Pinar del Río (Sinopsis de las aves fósiles de Cueva El Abrón, Pinar del Río, Cuba). *En* Biodiversidad Paleontológica del Archipiélago Cubano: Bases Cartográficas y Conservacionistas. Informe final del proyecto 022 AMA-CITMA; 074 MNHN. (CD ROM, inédito). 13 pp.

**PROBLEMA DE INVESTIGACIÓN.**

El conocimiento de la fauna de mamíferos extintos y vivientes del depósito "El Abrón", Sierra La Güira, Municipio Los Palacios, Pinar del Río, Cuba, está restringido solamente al análisis de uno de sus niveles estratigráficos (Nivel VII), el cual es el de mayor espesor, pero no el de mayor profundidad, ni antigüedad. La fauna estudiada de este importante nivel se limita a la composición taxonómica de aves fósiles (Suárez y Díaz-Franco, 2011; Suárez y Olson, 2015) y a la descripción de una nueva especie de murciélago para Cuba (Suárez y Díaz-Franco, 2003).

**HIPÓTESIS**

El estudio paleontológico del Nivel IX del depósito "El Abrón" permitirá incrementar la fauna de microvertebrados mamíferos reportada con anterioridad para el yacimiento y para la localidad geográfica en la que se encuentra ubicada.

**Objetivo general**

- Determinar la composición taxonómica de la fauna de mamíferos extinguidos, y vivientes, mediante el estudio paleontológico del nivel IX, depósito "El Abrón", Pinar del Río, Cuba.

**Objetivos específicos**

- Identificar los materiales paleontológicos pertenecientes a las especies de mamíferos extinguidos y vivientes.
- Clasificar el material osteológico correspondiente a la Clase Mammalia.
- Valorar el estado de conservación de los restos óseos y su relación con el origen del depósito.

# CAPÍTULO 1. REVISIÓN BIBLIOGRÁFICA

## 1.1. La importancia de los estudios paleontológicos

La Paleontología (del griego *palaios* = antiguo, *ontos* = ser, y *logía* = tratado, estudio) es una ciencia multidisciplinaria que se nutre del conocimiento de varias disciplinas científicas, esencialmente de la biología y la geología, basando su estudio en el conocimiento de la vida que habitó épocas geológicas pasadas (las paleobiotas) mediante los restos que han quedado en la corteza terrestre (los *fósiles*), permitiendo entender la actual composición (biodiversidad) y distribución (biogeografía) de los seres vivos sobre la Tierra (Meléndez, 1979; Aguirre, 1989; Meléndez, 1990; López-Martínez y Truyols-Santonja, 1994; Benton, 2005).

El estudio de los restos paleontológicos –fosilizados o no– permiten determinar el origen, evolución, filogenia, y causas de las extinciones de estas paleobiotas, así como los factores paleoecológicos, paleogeográficos, y paleoclimáticos que incidieron en ello, basándose en tres disciplinas esenciales: (1) la paleobiología, (2) la tafonomía, y (3) la biocronología (Fernández-López, 1989; López-Martínez y Truyols-Santonja, 1994; Fernández-López, 2000a; Benton, 2005).

Los autores citados anteriormente consideran que los estudios paleobiológicos se encargan específicamente del estudio de la sistemática, la anatomía comparada, la paleofisiología, entre otros, de los organismos del pasado, a los cuales se les denomina como entidades paleobiológicas, y que constituyen la fuente de información primaria de la paleontología: *los fósiles*.

La Tafonomía (del griego = *taphos*, enterramiento, y *nomos* = ley) estudia los procesos de fosilización y la formación de los yacimientos de fósiles, intentando descifrar la manera en que ocurrieron los procesos de acumulación que finalmente contribuyeron al origen de los depósitos, y los estados de conservación de los materiales que en ellos se han acumulado a lo largo del tiempo (Fernández-López, 1986, 1999, 2000a; 2000b; 2005). También el estudio de los factores que han sido responsables de la conservación de los restos fósiles, tales como estadios de meteorización (Behrensmeyer, 1978; Fernández-Jalvo *et al.,* 2002), es objeto de estudio de esta disciplina.

Respecto a la Biocronología plantean que esta disciplina intenta descifrar, mediante los fósiles, la cronología de las especies extinguidas, además de su organización secuencial en el tiempo y la cronología de paleobioeventos asociados al proceso de extinción de muchas de las especies que aparecen en el registro estratigráfico. No obstante, algunos autores consideran que la estimación de la edad biológica (ontogénica) de los individuos actuales o pretéritos puede ser denominada como Biocronología Biológica y no debe ser confundida con la bioestratinomía, que se restringe solamente a datar los cuerpos rocosos que contienen los fósiles y no los fósiles específicamente (Fernández-López, 1991; López-Martínez y Truyols, 1994, Fernández-López y Fernández-Jalvo, 2002).

Específicamente, los restos de las especies extintas que aparecen en el registro estratigráfico, se encuentren fosilizados o en vía de fosilización, se denominan *fósiles,* los cuales permanecen conservados en determinadas estructuras geológicas que actúan como reservorio de los mismos y denominados yacimientos o depósitos paleontológicos (Fernández-López, 2000b; 2000c).

La naturaleza de los fósiles responde a diferentes tipos de materiales, tales como: restos óseos, dientes, huellas, réplicas, entre otros, y en su interacción con la biosfera y la litosfera sufren transformaciones generadas por estos ambientes convirtiéndolos en una valiosa fuente de información sobre la vida pasada en la Tierra (Simpson, 1985; Fernández-López, 1989; Bates y Jackson, 1987; Bromley, 1990; Doménech y Martinell, 1996; Fernández-López, 2000b; Fernández-Jalvo *et al.,* 2002; Lacasa, 2010, entre otros).

### 1.2. Los yacimientos paleontológicos y su tipificación

Los yacimientos paleontológicos se reconocen como aquellas estructuras de origen geológico que presentan determinadas condiciones y características para la captación y conservación de restos paleobiológicos, los cuales, tras el paso del tiempo, pueden llegar a transformarse en fósiles o asociaciones de fósiles (Fernández-López, 1984; 1999; 2000a). La capacidad de almacenar y conservar estos restos no está relacionada solamente con las condiciones geomorfológicas del depósito sino también con los procesos naturales (paleoclimáticos fundamentalmente) que incidieron en su evolución durante muchos años. Generalmente los restos conservados en los yacimientos paleontológicos reflejan algunas huellas de estos eventos pasados, y la paleontología mediante la disciplina tafonomía intenta interpretarlos (Fernández-López, 1989; López-Martínez y Truyols-Santonja, 1994; Fernández-López, 1986, 1999, 2000b; 2000c).

Algunos autores han clasificado o descrito las condiciones de los yacimientos con la intención de establecer tipificaciones acorde al origen de los materiales y las características de los sedimentos acumulados, tal como puede apreciarse en los trabajos de Raup y Stanley [1978]; López-Martínez y Truyols-Santonja (1994), Fernández-López (2000c), Gruber (2007), Quinif (2009), entre otros. En estos trabajos se pueden encontrar referencias a yacimientos o depósitos de origen sedimentario con características relacionadas al medio de sedimentación (marino, continental de tipo fluvial o lacustre, o de transición). También existen clasificaciones atendiendo al desarrollo cársico y el comportamiento de la circulación de las aguas subterráneas y superficiales que condiciona la manera en que se transportan, conservan y acumulan los restos en un yacimiento, tal como se evidencia en el trabajo de Quinif (2009).

Para el caso del sustrato geológico cubano, 67831 $km^2$ son rocas carsificadas, lo que representa 66% respecto a los 110,992 $km^2$ que conforman el territorio nacional (Iturralde-Vinent y Gutiérrez- Doménech, 1999). Según estos autores la combinación de factores topográficos, litológicos, hidrodinámicos, la acción química del agua, entre otros, con la humedad del clima tropical, condicionan la variedad de formaciones cársicas en Cuba (véase también Skwaletski e Iturralde-Vinent, 1971; Iturralde-Vinent, 1972; Núñez *et al.,* 1988). La amplia variedad de fenómenos cársicos presentes en Cuba (cuevas, sumideros, fisuras, poljas) son reservorios de la mayoría de los materiales paleontológicos correspondientes al Pleistoceno-Holoceno, tal como se refleja en los trabajos de Arredondo (1970), Silva (1974), Woloszyn y Silva (1977), Acevedo y Arredondo (1982); Silva *et al.* [2008].

Evidentemente, ante tal variedad, es de suponerse condiciones de conservación específicas para los materiales acumulados, por lo que definir o intentar establecer clasificaciones de yacimientos en coherente relación con la manera en que se acumularon y conservaron los materiales es de extrema importancia para la paleontología (Woloszyn y Silva, 1977; Acevedo y Arredondo, 1982).

En el caso del trabajo de Woloszyn y Silva (1977) se proponen cuatro tipos de depósitos para explicar la acumulación de restos de vertebrados cubanos en los yacimientos fosilíferos.

El clasificado como **Tipo A** es caracterizado como aquellos originados por la depredación de la lechuza *(Tyto)*, las cuales forman sus nidos a las entradas de las cuevas. Los restos alimentarios de su actividad trófica se acumulan primeramente debajo de los nidos y luego pueden ser transportados por el agua y depositados en el interior de las cuevas. La composición de estos depósitos refleja las preferencias alimentarias de la lechuza: especies relativamente pequeñas de mamíferos, aves, reptiles y anfibios. Estos depósitos por su cercanía a las entradas están expuestos a diversos agentes destructivos, y por tanto los depósitos de este tipo en la actualidad no deben tener mayor antigüedad. El **Tipo B** responde a depósitos formados en grandes grietas, sumideros, o casimbas, que actúan como embudos de captación de las aguas circundantes. En este caso se depositaron en el sumidero, arrastrados por las aguas, los restos de animales que murieron fuera. Las cuevas con mucha frecuencia representan total o parcialmente este tipo de sumidero. En los casos en que se trata de rellenos de grietas o sumideros expuestos por la explotación de una cantera, los depósitos pueden ser particularmente antiguos. El **Tipo C** está relacionado con la actividad antrópica precolombina, es decir, entierros cavernarios aborígenes que presentan restos de animales incorporados al depósito en calidad de ofrenda ritual, o también a los depósitos constituidos por residuos de la dieta de los propios aborígenes o restos zooarqueológicos. Estos restos aparecen comúnmente en las inmediaciones de las entradas de las cuevas y carecen de mayor antigüedad. Finalmente, el **Tipo D** se corresponde con los depósitos fosilíferos en el interior de las cuevas, constituidos casi exclusivamente por restos de murciélagos y representan el resultado de la muerte súbita de la o las diferentes especies que ocupaban la cueva simultáneamente debido al evento responsable de ello. Estos depósitos son más antiguos que los tipos A y C. Se presentan en las zonas más internas de las cuevas.

Por otra parte la propuesta de Acevedo y Arredondo (1982) adiciona cinco nuevos tipos a la clasificación de Woloszyn y Silva (1977), siguiendo el mismo criterio y poniendo de ejemplo localidades supuestamente representativas de cada tipo, donde el **Tipo E** serían depósitos producidos por inundaciones fluviales y originados al aire libre; el **Tipo F** originados al aire libre y actualmente contenidos en suelos fósiles; el **Tipo G** es el resultado de acumulaciones superficiales de restos no enterrados producto de caída de animales vivos en cavidades y pérdidas en las cavernas; el **Tipo H** que se refiere a acumulaciones en manantiales, fuentes o depósitos de agua donde los animales murieron ahogados o producto de la actividad trófica de cocodrilos; y finalmente el **Tipo I** que incluye solamente a aquellos restos acumulados en depósitos asfaltíferos.

### 1.3. El registro paleontológico de vertebrados cubanos

Desde el punto de vista geológico el archipiélago cubano es extremadamente joven comparado con otras áreas del planeta, tal como puede corroborarse en el trabajo de Iturralde-Vinent y MacPhee (1999). Sin embargo, cuando se aborda la temática del registro paleontológico cubano se considera en general que abarca desde el período Jurásico tardío cuando comienzan los primeros vestigios de la región caribeña (protocaribe), tal como se refleja en el trabajo de Iturralde-Vinent (2003). Así mismo en el trabajo de Rojas-Consuegra (2013), se ilustra una columna del registro macrofósil de Cuba que abarca también desde el Jurásico tardío hasta el Cuaternario, teniendo en cuenta fundamentalmente grupos zoológicos como invertebrados y vertebrados.

El "tardío" inicio de la formación de nuestro archipiélago, entre 45-35 MaAP, según Iturralde-Vinent y MacPhee (1999), limita el registro en general entre estos dos períodos. No obstante, estudios realizados por (Iturralde-Vinent, 2003) aportan datos acerca de microfósiles foraminíferos (Schwagerínidos) de edad Paleozoica, entre 540-250 MaAP. Para el registro de macrofósiles los más antiguos conocidos del territorio cubano, son del Jurásico Inferior a Medio (Helechos y Moluscos), que tienen menos de 200 MaAP, corroborando la tardía formación de nuestro archipiélago.

En el trabajo de Rojas-Consuegra (2013) se refleja claramente el incremento de taxones fósiles descritos para la ciencia cubana entre los períodos Neógeno y Cuaternario. Específicamente para el grupo de vertebrados fósiles cubanos las épocas Pleistoceno-Holoceno son las de mayor cantidad de depósitos conocidos, y aún más si se trata de mamíferos (Silva *et al.,* [2008]; Arredondo, 2011a); por lo que ha tenido una mejor atención por parte de los paleontólogos cubanos y una gran cantidad de estudios, aunque no en la misma medida para todos los grupos zoológicos. Esto se debe fundamentalmente a la mejor accesibilidad a los depósitos que se originaron en ambos períodos y sus estados de conservación.

Es importante resaltar que durante el período Pleistoceno la paleogeografía del archipiélago cubano mostró espacios terrestres muy reducidos con respecto a otras etapas dentro del mismo período (Iturralde-Vinent y MacPhee, 1999; 2004), favoreciendo los procesos de extinción de la fauna terrestre fundamentalmente. También hay que tener en cuenta que la mayoría de los depósitos paleontológicos actuales contentivos de vertebrados son el resultado de acumulaciones producto de la acción de arrastre de las aguas superficiales, y en menor medida de las aguas subterráneas, las cuales acarrearon los restos expuestos en ambientes pasados hacia espacios (fundamentalmente cársicos) aptos para su recepción (Woloszyn y Silva, 1977; Acevedo y Arredondo, 1982), donde los materiales de mayor talla y resistencia ante los factores de alteración (o tafonómicos) persistieron en el tiempo.

Por tal razón los trabajos paleontológicos cubanos orientados hacia el grupo de los vertebrados refleja un mejor desarrollo para dos grupos esencialmente: las aves, y en mayor medida, los mamíferos. Para ambos casos muchísimos son los trabajos que han aportado nuevos registros de especies, análisis biogeográficos, entre otros aspectos; sin embargo el conocimiento más actualizado al respecto se resume en los trabajos de Suárez (2005); Suárez y Olson (2007); Olson y Suárez (2008a; 2008b; 2008c); Suárez y Olson (2009; 2015), para el caso de las aves, y Silva *et al.* [2008], y Borroto-Páez y Mancina (2011 *eds*.) para el caso de los mamíferos.

Ambos grupos taxonómicos poseen la ventaja de tener especies de gran talla ósea, aspecto este que favorece la conservación de sus restos en algunos espacios que ocuparon y quedaron registrados. En el caso de las aves, si bien sus huesos son frágiles, algunas de las especies de este grupo pueden conocerse hoy gracias a la posibilidad que brindaron los nichos que ocuparon de mantener conservados sus restos, lejos de la acción meteórica a la que estuvieron expuestos muchos otros materiales óseos de vertebrados, y que se hallan en menor grado de conservación.

Para el caso de otros grupos de vertebrados, como los anfibios y los reptiles, aún carecen de suficiente atención por parte de los paleontólogos, aspecto este ligado a la carencia de materiales osteológicos de referencia y de los estados de conservación de los materiales óseos de las especies que los componen (Alemany *et al.,* 2015). Los ambientes en que habitaron las especies de ambas clases no son tan favorables para la conservación de sus restos en el registro paleontológico, con la excepción de los cocodrilos, los cuales aparecen con frecuencia en determinados depósitos.

Atendiendo a los trabajos anteriormente mencionados podemos decir que dentro de los materiales más comunes y mejor estudiados que se encuentran en el registro fósil cubano se destacan los restos óseos y dientes de mamíferos tales como el Orden Pilosa (perezosos), Orden Primates (monos), Orden Rodentia (jutías y ratas espinosas), Orden Soricomorpha (musarañas), y el Orden Chiroptera (murciélagos), y los órdenes Carnívora (Perros y Focas), Cetácea (Ballenas) y Sirenia (Manatíes y Dugónidos), en menor medida. Para el caso de las aves, son las depredadoras y de hábitos nocturnos las más comunes dentro los órdenes Falconiformes y Strigiformes respectivamente. Para el caso de los reptiles el Suborden Sauria (Género *Anolis* fundamentalmente) y el Orden Squamata tienen representantes dentro del registro fósil; mientras que para los anfibios es el Orden Anura el único representante.

### 1.4. Algunas consideraciones acerca de los mamíferos fósiles de Cuba

Como ya mencionamos con anterioridad, durante el intervalo geológico comprendido entre entre 45-35 MaAP tiene sus inicios la geografía de nuestro archipiélago (Iturralde-Vinent y MacPhee, 1999) y por ende la fauna que lo compone. Diversas han sido las hipótesis que se han planteado sobre el origen de las Antillas y su biota en las últimas décadas (Rosen, 1975; Hedges *et al.,* 1992; Hedges, 1996a; 1996b; Iturralde-Vinent y MacPhee, 1999; Iturralde-Vinent, 2003; 2005a; 2005b; Hedges, 2006, entre otros). Este último autor resume la problemática hasta el momento conocida acerca de las variantes para el poblamiento de las Antillas por la fauna ancestral, tales como la dispersión y la vicarianza, dejando claro que todavía la ciencia no está en condiciones de desestimar alguna de ellas.

Por otra parte, según Silva *et al.* [2008], un tema de particular importancia lo constituye el análisis de la fauna fósil terrestre no voladora (debido al sinnúmero de interrogantes que se generan ante cualquier análisis hecho hasta el momento, convirtiendo este tema en un espacio académico de grandes discusiones en la actualidad (Alí, 2012).

Para el caso particular de los mamíferos se sugiere la hipótesis de que algunos grupos ancestrales pudieron haber empezado a colonizar nuestro archipiélago alrededor de 40 MaAP, durante el período Paleógeno (épocas Eoceno medio e inferior y Oligoceno), teniendo en cuenta el escenario paleogeográfico del Caribe para ese entonces expuesto en el trabajo de Iturralde-Vinent y MacPhee (1999), donde algunas porciones de tierras emergidas pudieron facilitar el poblamiento de las especies que se conocen hoy, con excepción de las introducidas antrópicamente (Silva *et al.* [2008]; Borroto-Páez y Mancina, 2011 *eds.*).

Sin embargo la evidencia fósil de las especies extintas más antiguas hasta el momento marcan geológicamente la antigüedad de nuestra fauna entre 25-17.5 MaAP [períodos Paleógeno (finales del Oligoceno) – Neógeno (gran parte del Mioceno)]. En los trabajos de MacPhee e Iturralde-Vinent (1994); y MacPhee *et al.* (2003) se describen nuevas especies de primates (*Paralouatta marianae*), perezosos (*Imagocnus zazae*), y roedores (*Zazamys veronicae*) siendo la localidad de Domo de Zaza, en la provincia de Sancti Spíritus, la localidad tipo de las mismas.

Independientemente de los múltiples trabajos relacionados a la paleontología de mamíferos cubanos el conocimiento sistemático y biogeográfico más actualizado al respecto, con la excepción de murciélagos y mamíferos marinos, se resume en el trabajo de Silva *et al.,* [2008]. Posteriormente, en la compilación realizada por Borroto-Páez y Mancina (2011, *eds.*) acerca de “Los Mamíferos en Cuba” se reflejan trabajos acerca de los grupos terrestres ya tratados en Silva *et al.* [2008], en los acápites desarrollados por Arredondo (2011a; 2011b; 2011c); Condis (2011); y Jiménez-Vázquez (2011), pero además en el acápite de Arredondo (2011a) se incluye la fauna de mamíferos marina no contenida en el trabajo de Silva *et al.* [2008], así como una actualización del conocimiento sobre los murciélagos extintos de Cuba desarrollada por Balseiro (2011) en otro acápite independiente. Finalmente Jiménez-Vázquez y Arredondo (2011) hacen alusión a la relación de los mamíferos cubanos y la arqueozoología, aspecto este no desligado de la presencia de algunos grupos de mamíferos en Cuba y también de la extinción de otros.

Atendiendo al conocimiento resumido en los trabajos mencionados anteriormente la presencia de mamíferos fósiles en el registro estratigráfico cubano responde a factores naturales y antropogénicos, donde algunos grupos son de reciente extinción, y la presencia de otros es consecuencia de la introducción de los mismos por comunidades aborígenes en períodos previos a la colonización.

Por otra parte la mayoría del registro de mamíferos fósiles de Cuba se concentra en depósitos contextualmente ubicados en el período Cuaternario, entre las épocas Pleistoceno (1,8 MaAP-12 kaAP) - Holoceno (12 kaAP - actualidad) lo que sugiere dos factores esenciales: (1) son los depósitos fosilíferos mejores conservados, lo que proporcionalmente favorece la conservación de los restos, y además, son geológicamente menos complejos de estudiar debido al tipo de formaciones que los caracterizan (cuevas, solapas, grietas cársicas, entre otras); y (2) durante este período y las épocas mencionadas han ocurrido eventos significativos de extinción, asociados a causas naturales en mayor medida Iturralde-Vinent y MacPhee (2004), y en menor medida las antropogénicas .

En relación a ambos factores (excluyendo las causas antropogénicas) se conoce que durante el Pleistoceno Superior el panorama paleogeográfico del archipiélago cubano mostró etapas donde los espacios terrestres quedaron muy reducidos con respecto a etapas anteriores, lo que debió favorecer considerablemente los procesos de extinción de la fauna fundamentalmente terrestre (Iturralde-Vinent y MacPhee, 1999; MacPhee e Iturralde-Vinent, 2000; Silva *et al.* [2008]).

Los factores geoclimáticos también incidieron fuertemente en el proceso de extinción de la megafauna cubana durante los períodos antes mencionados. Dentro de estos cambios climáticos, las variaciones del nivel del mar han jugado un papel esencial, tanto en la modificación de las líneas de costa y cantidad de terrenos emergidos, así como en el tipo de vegetación y condiciones ecológicas en el área caribeña en general como en el territorio cubano (Pregill y Olson, 1981; Ortega y Arcia, 1982; Iturralde-Vinent, 2003; 2005a; 2005b;

Peros *et al.,* 2007; entre otros). El aumento de los procesos hidrológicos como resultado de estos cambios climáticos incidió en el origen de muchos depósitos y su acumulación actual.

Todo lo anteriormente expuesto está en total correspondencia con el hecho de que, si bien las Antillas no ocupan una gran parte dentro de la superficie total del planeta, los estudios paleontológicos de vertebrados apuntan a que esta región presenta una alta tasa de extinción para los mamíferos. Mancina y Borroto (2011:20) señalan que *"De las 56 especies de mamíferos terrestres que han habitado la isla de Cuba, 41 % se encuentra extinto."*; donde algunos órdenes como Pilosa y Primates no presentan especies vivas, es decir, se encuentran totalmente extintos (ver MacPhee e Iturralde-Vinent, 1994; Horovitz y MacPhee, 1999; MacPhee *et al.,* 2003; Silva *et al.* [2008]).

### 1.5. La lechuza (*Tyto*) como agente natural de acumulación de restos óseos

Son los depósitos fosilíferos cubanos de edad Cuaternario los que fundamentalmente contienen los registros de las aves fósiles, generalmente en aquellos comprendidos en la época geológica Pleistoceno Superior (Acevedo y Arredondo 1982, Arredondo 1984). Los estudios de los restos de las regurgitaciones (*egagrópilas*) que realizaron estas aves durante su acción trófica están considerados de gran importancia, tanto para estudios taxonómicos y ecológicos actuales, como paleontológicos. Algunos trabajos en este sentido pueden mencionarse tales como los de Arredondo (1982), Jiménez-Vázquez *et al.* (2005), Hernández-Muñoz y Mancina (2011); López-Ricardo y Borroto-Páez (2012), Suárez y Olson (2015), entre otros.

Dentro de las aves en general, el orden Strigiformes (rapaces nocturnas) es el segundo de mayor número de especies extintas para Cuba (8 especies) solamente antecedido por el orden Falconiformes (rapaces diurnas) que está compuesto por 10 especies (Suárez, 2005). Según este autor este grupo de aves rapaces nocturnas sufrieron 53,3% de extinción, donde la familia Strigidae perdió completamente uno de sus dos géneros endémicos (*Ornimegalonyx*), además de 8 especies endémicas, de un total de 10.

Las especies que componen el género *Tyto* (Familia Tytonidae), conocidas como lechuzas, inciden frecuentemente en la formación de residuarios de *egagrópilas*, y por ende están entre las mayormente estudiadas paleontológicamente. Dentro de las especies que se conocen hasta el momento dentro de este género se encuentra *T. noeli*, *T. pollens* (=*T. riveroi,* sp. sinonimizada en Suárez y Olson, 2015)*,* ambas extintas, y *T. alba (furcata),* actualmente presente en nuestro archipiélago y responsable de la mayor cantidad de residuarios paleontológicos dentro de las últimas etapas de la época Holoceno, además, está considerada una de las aves con mayor distribución a nivel mundial al estar representada en cinco continentes (López-Ricardo y Borroto-Páez, 2012). Estas rapaces son usualmente clasificadas como nocturnas, naturalmente debido a sus hábitos, los cuales han sido estudiados por diversos autores en todo el mundo (Martínez y López, 1999; Ramírez *et al.,* 2000; Pardiñas y Cirignoli, 2002; Aliaga-Rossel y Tarifa, 2005; Stangl y Shipley, 2005; Leonardi y Dell´Arte, 2006; Cenizo y Reyes, 2008).

El estudio de yacimientos fosilíferos cubanos producidos por la acumulación de estrigiformes no ha sido muy abundante, por lo que este tipo de estudio es un tema poco tratado para nuestro archipiélago, en consideración a las referencias actuales. Algunos autores destacan que este tipo de acumulación en ambientes cársicos se encuentra entre los más frecuentes en correspondencia con la naturaleza geológica del país, aspecto mencionado en el trabajo de Woloszyn y Silva (1977).

Tanto en el trabajo de Jiménez-Vázquez *et al.* (2005), como en el de López-Ricardo y Borroto-Páez (2012), se destacan algunas características generales de la composición del contenido de las regurgitaciones (*egagrópilas*) de las aves estrígidas responsables de la formación de los depósitos de este tipo. Se plantea que son las especies de menor tamaño dentro de los vertebrados en general las que mayormente forman parte de la dieta de estas aves, donde naturalmente, el análisis de la fauna de micromamíferos ha sido la mejor estudiada.

Se conoce que la selección de la dieta está determinada por la disponibilidad y vulnerabilidad de las presas en el hábitat. Este planteamiento puede ser corroborado en los trabajos de Glue (1974), Marti (1988), Bellocq y Kravetz (1993), Martínez y López (1999), Bellocq (2000), Ramírez *et al.,* (2000), Andrade *et al*. (2002), Pardiñas y Cirignoli (2002), Álvarez-Castañeda *et al*. (2004), Stangl y Shipley (2005), Aliaga-Rossel y Tarifa (2005), Begall (2005), Leonardi y Dell´Arte (2006), Velarde *et al.* (2007), Delgado y Calderón (2007), Cenizo y Reyes (2008), Fernández *et al*. (2009), Hernández-Muñoz y Mancina (2011), Fuentes *et al*. (2012), López-Ricardo y Borroto-Páez (2012). Dentro de los mamíferos (micromamíferos) son los roedores en general la base fundamental de la dieta de lechuzas actuales, aspecto este con evidente conexión en estudios de residuarios paleornitológicos (Jiménez-Vázquez *et al.*, 2005).

Un aspecto a tener en cuenta en relación con la importancia de estudiar la composición de las *egagrópilas* es detectar la presencia de fauna de vertebrados pequeños, tanto en el sentido de estudiar su distribución, ecología en general, etc., así como intentar encontrar restos de especies reportadas como extintas. Este último propósito se debe a que solamente las lechuzas pueden acceder a determinados sectores de la geografía donde algunas especies de vertebrados pequeños pueden haber sobrevivido, tanto a los cambios ambientales a lo largo de períodos geológicos como a los efectos antropogénicos de los últimos miles de años.

Un ejemplo de extrema importancia para la paleontología cubana es el caso de la especie *Nesophontes micrus* dentro del orden Soricomorpha (Familia Nesophontidae). Abundantes son los restos de esta especie en cuevas (fundamentalmente) y en solapas cársicas, aspecto este relacionado directamente desde el pasado siglo con la acción depredadora de *Tyto alba* (Aguayo y Howell, 1954), aunque se conoce en posteriores estudios (Suárez y Díaz-Franco, 2003) que la especie *Tyto noeli* contribuyó milenios antes a la acumulación de fósiles en ambientes cársicos.

Se considera que el género *Tyto* presenta al menos una especie fósil más (Suárez y Díaz-Franco, 2003) la cual aún se encuentra en proceso de estudio y publicación. Los restos de esta posible especie fueron hallados en el depósito objeto de estudio, lo que incrementa su importancia, y además, este aspecto puede tenerse en cuenta a la hora de responsabilizar al género *Tyto* de la acumulación (casi total) de los restos estudiados, así como de algunos aspectos tafonómicos relacionados con la conservación de los restos, sobre todo el grado de fragmentación.

Finalmente, el estudio de las acumulaciones generadas por *Tyto* durante diversos períodos y épocas geológicas representa gran importancia para conocer mediante los restos (fósiles o subfósiles) la composición de la fauna de mamíferos del pasado.

### 1.6. El depósito paleontológico "El Abrón"

El único trabajo publicado que recoge los resultados científicos de estudios paleontológicos llevadas a cabo en el depósito fosilífero objeto de estudio es el de Suárez y Díaz-Franco (2003). En esta importante publicación se analiza el material de uno de los niveles trabajados en el depósito, en el cual se describe una nueva especie de murciélago para Cuba: *Phyllops silvai*, además de que reporta la presencia de millones de restos óseos acumulados de microvertebrados, donde en las capas más superficiales (o tardías) se aprecia la incorporación de taxones introducidos (género *Rattus*). Este trabajo constituye el antecedente principal de cualquier investigación que se realice posteriormente en el yacimiento, aunque su nivel de especialización en la temática abordada no lo convierte en un trabajo de referencia total acerca de la fauna hallada en el depósito. No obstante, en el trabajo citado se hace alusión al tipo de yacimientos según su génesis teniendo en cuenta los criterios de Woloszyn y Silva (1977).

En esta publicación (Suárez y Díaz-Franco, 2003) se presenta además un fechado radio carbónico realizado sobre material óseo de la especie extinta de lechuza *Tyto noeli* que ubica su acción trófica alrededor de 17 406 ±161 años AP, y se alerta sobre la presencia de una nueva especie de este mismo género. La importancia de este fechado, y por ende, una más del trabajo, es que el nivel estudiado (VII) es el penúltimo en la estratigrafía, y es el de mayor espesor. Aunque no se conoce exactamente el punto en el estrato donde fueron tomados los restos a fechar, este segmento termina en los 2,00 m de profundidad, a 60 cm de donde comenzaría el nivel objeto de estudio de este trabajo (IX), lo que sugiere, en principio, una mayor antigüedad para los restos depositados tanto en este nivel como en el anterior (VIII). Otro aspecto a considerar es que este fechado constituye un punto de referencia importante a la hora de establecer relaciones paleoecológicas tanto para la especie fechada como para sus presas. Sería posible a partir de la información contextual que ofrece relacionarlos con trabajos de tipo paleopalinológicos y paleoclimáticos, en aras de reconstruir el segmento de la historia natural que ocuparon las especies relacionadas.

Posteriormente al trabajo antes mencionado, en informes de investigación, fueron tratados algunos de los resultados más relevantes del estudio de este depósito. Se describieron ocho niveles estratigráficos considerados de origen natural, reflejados inicialmente en el trabajo de Suárez y Díaz-Franco (2011:3-4), aunque en 2015, a partir del análisis de los trabajos realizados anteriormente, se infieren determinadas características sedimentarias del último estrato [véase Nivel IX (2,60 – 2,90 m)]. Los niveles hasta ahora considerados se resumen a continuación, y una reconstrucción gráfica de los mismos (hasta el Nivel VIII) puede consultarse en la Fig. 1.

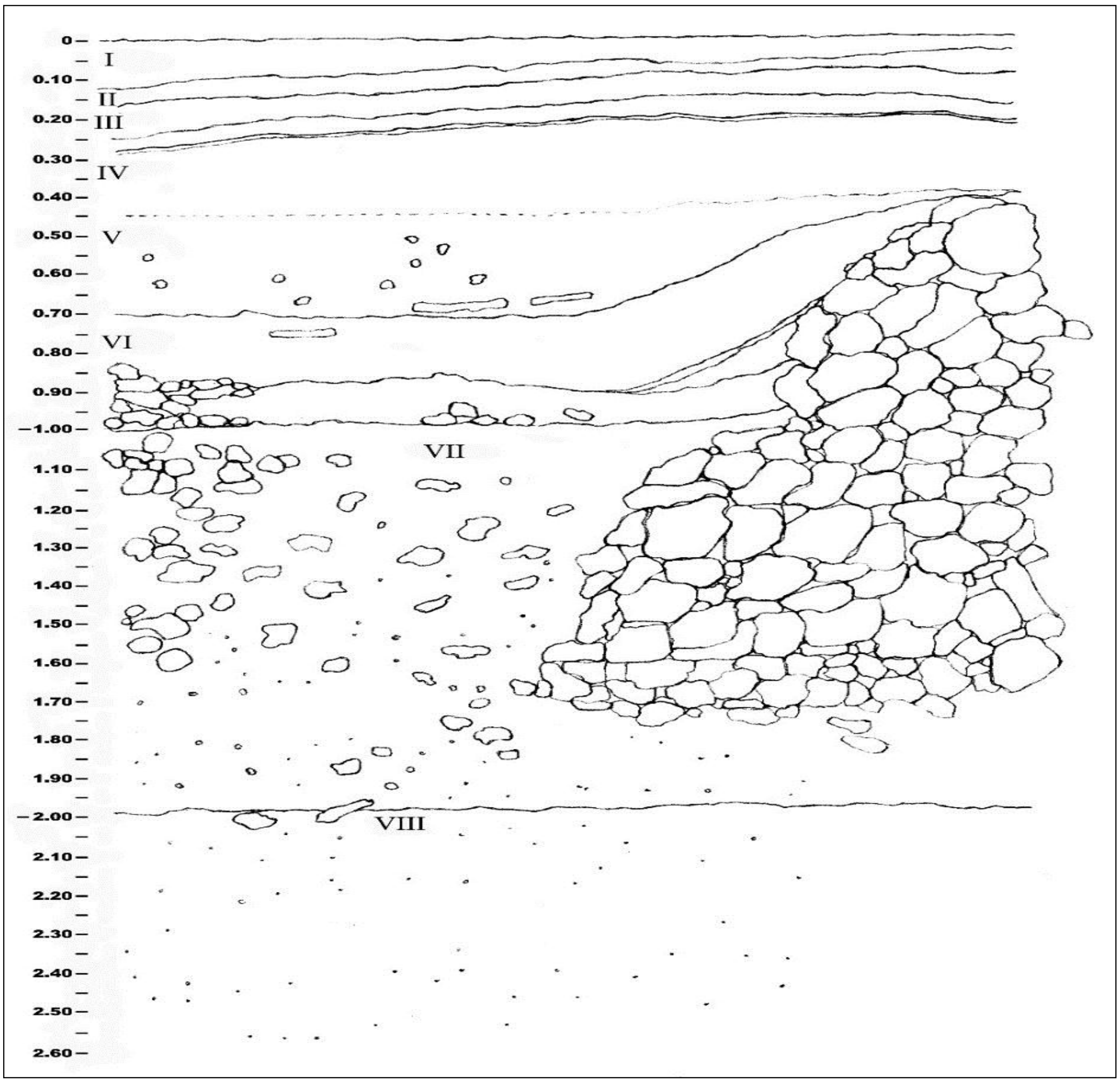

**FIGURA 1.** Estratigrafía del depósito paleontológico "EL Abrón". Nótese la ausencia de la delimitación estratigráfica del Nivel IX.

***Nivel I (0,00 – 0,12 m):*** *Color pardo, suelto, friable, con pocos fragmentos de rocas. Muy rico en restos de vertebrados fósiles y actuales.*

***Nivel II (0,12 – 0,18 m):*** *Color grisáceo, con fragmentos de rocas pequeñas. El color cambia, descendentemente, de gris con parches claros y cremas a pardo. Muy rico en restos de vertebrados, la mayoría de color negro.*

***Nivel III (0,18 – 0,28 m):*** *Color pardo a cremoso. Muy rica en fósiles.*

***Nivel IV (0,28 – 0,45 m):*** *Color pardo, ligeramente más oscuro que el del nivel siguiente. La parte superior del nivel IV está marcada por una fina capa parda oscura. Se observan muy pocos fósiles.*

***Nivel V (0,45 – 0,70 m):*** *Color pardo, a mayor profundidad comienza a observarse un incremento del carbonato precipitado, y empiezan a aparecer unas estructuras semejantes a nódulos de caliza a partir de los 0,50 m hasta 0,70 m. Se colectó madera entre 0,65 y 0,70 m. Hay muy pocos fósiles.*

***Nivel VI (0,70 – 1,00 m):*** *Color crema muy claro (tal vez por un alto contenido de carbonatos), hay rocas relativamente grandes. Se colectó madera entre 0,75 y 0,80 m. Este nivel es particularmente rico en vertebrados; aunque variable, es el de mayor espesor (0,20 m máximo), forma toda una capa continua de restos óseos.*

***Nivel VII (1,00 – 2,00 m):*** *Color pardo rojizo, quizás por la humedad se observa amarillento (tal vez seco sea más claro). En todo el perfil (desde 1,40 – 1,72 m), aparecen abundantes fragmentos de roca de grandes dimensiones relacionados con un evento de desplome. En la parte basal del nivel se aprecia un acuñamiento de fragmentos rocosos que debe corresponder al comienzo del evento de derrumbe. La riqueza de vertebrados disminuye con la profundidad.*

***Nivel VIII (2,00 m – 2,60 m):*** *Color rojizo a pardo, más oscuro que el del nivel superior; hay muy pocos fragmentos de roca. Los sedimentos se extienden por debajo del acuñamiento de rocas (que termina a 1,72 m de profundidad con respecto a la superficie). La base del acuñamiento debe corresponder al antiguo nivel del piso de la cavidad. Se recuperan muy pocos fósiles en general, aunque siguen apareciendo restos de vertebrados y moluscos. Entre 2,25 – 2,30 m aparece una fina capa de fósiles, originando cierto enriquecimiento local. A 2,50 m ya se recuperan muy pocos fósiles del tamiz.*

***Nivel IX (2,60 – 2,90 m):*** *Según Rojas-Consuegra (2015, com. pers.), este segmento de la estratigrafía del depósito presenta una coloración de tipo rojizo a pardo, más claro de color gris amarillento cuando está seco, debido también al contenido de carbonato, que forma una fina capilla sobre los restos. Después del lavado con agua se observan elementos biogénicos de diferentes coloraciones: rojizo (predominantes), grisáceos (abundantes, aún recubiertos) y gris oscuro (solo algunos). Contiene fragmentos sólidos, posiblemente de carbonato, de calcita cristalina y arcilla-arenosa algo compacta, que alcanzan hasta 2 ó 3 cm, además, aparecen otros clastos pequeños de variadas composiciones. El material fósil se calcula que representa un 10 – 15 % del volumen total del sedimento muestreado.*

Un aspecto de relevante importancia que es mencionado en este trabajo es acerca del ambiente *diagenético* del contexto en que se encontraron los restos óseos, el cual, según estos autores, favoreció la preservación de abundantes materiales paleontológicos de diferentes grupos de vertebrados, los cuales aparecen altamente mineralizados y coloreados en relación con el color de las capas en que se encuentran. También hacen alusión a los grados de dispersión (cualitativamente) de los restos óseos destacando la no concentración de los mismos en un área en particular, lo que los hace inferir en la posibilidad de algún mecanismo de transporte posterior a la acumulación inicial y disgregación de las *egagrópilas*.

Finalmente, y con carácter reciente, en el trabajo de Suárez y Olson (2015) se incluyen datos de este depósito paleontológico cuando se refiere a una de las dos especies reportadas para el mismo (*Tyto noeli*). Estos autores destacan que restos de esta especie aparecen mezclados con restos juveniles de sus principales presas (micromamíferos), tales como *Geocapromys columbianus* y *Boromys offella* (ambos géneros extintos dentro del orden Rodentia). Es importante resaltar que en los trabajos anteriores (Suárez y Díaz-Franco, 2003; Silva *et al.* [2008]; Suárez y Díaz-Franco, 2011) no se había reportado la especie del género *Geocapromys* (*G. columbianus*); el género *Boromys* (*B. offella y B. torrei*) si había sido reportado en Silva *et al.* [2008], incluso directamente del depósito según el catálogo de localidades que ofrecen estos autores. Por tanto, en este trabajo (Suárez y Olson, 2015), aunque no se refleje como nuevo reporte, se hace mención por vez primera a la presencia del género *Geocapromys*, con su única especie representada para Cuba, en la acumulación de *egagrópilas* del depósito objeto de estudio.

## CAPÍTILO 2. MATERIALES Y MÉTODOS

### 2.1. Área de estudio

El área objeto de estudio se encuentra en la Cordillera de Guaniguanico que es un sistema montañoso que abarca varios municipios de la provincia Pinar del Río dentro de los cuales se encuentra Los Palacios (Comisión Nacional de Nombres Geográficos, 2000). Según este mismo autor se pueden apreciar dos grupos de paisajes con diferentes características litológicas, denominados como la Sierra de los Órganos y la Sierra del Rosario. En la Sierra de los Órganos, específicamente en la porción geográfica correspondiente al municipio Los Palacios, se encuentra la Sierra de la Güira (Fig. 2), entre los 22°38´ y 22°40´, latitud norte y los 83°23´ y 83°27´, longitud oeste y abarca un área de aproximadamente 12,52 km$^2$, con una altura máxima de 514 m s.n.m (véase Regalado y Lóriga, 2010).

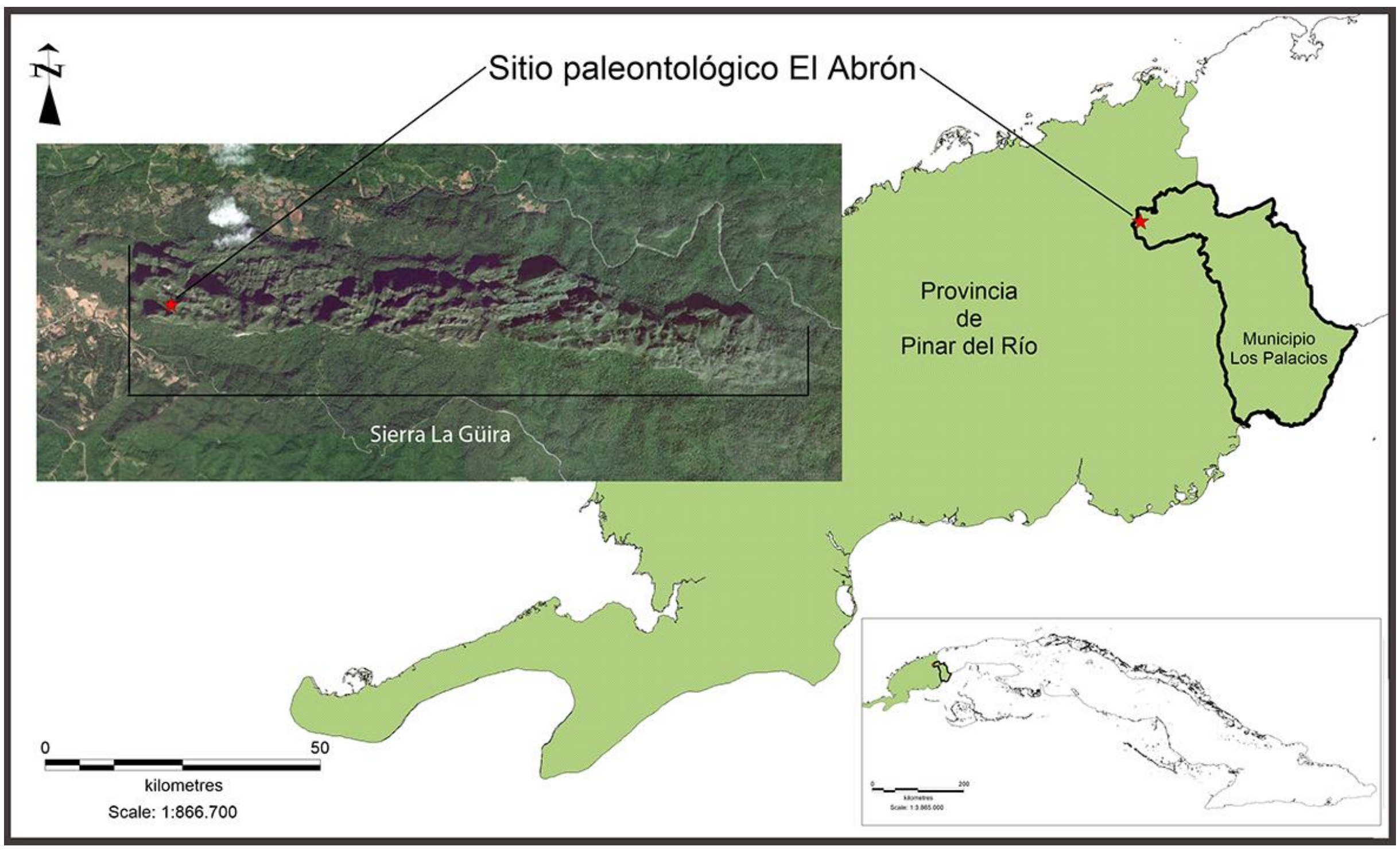

**FIGURA 2**. Representación del área de estudio. La estrella roja indica la ubicación geográfica del depósito "El Abrón", Sierra La Güira, Los Palacios, Pinar del Río, Cuba.

Según datos reflejados en Comisión Nacional de Nombres Geográficos (2000) esta sierra está básicamente constituida por rocas calizas, lutitas y argilitas del Jurásico Superior; areniscas, esquistos y pedernales del Paleoceno – Eoceno. Este autor agrega que el relieve está formado por montañas tectónico-litológicas pequeñas en forma de bloque; que los suelos son ferralítico amarillento, fersialítico pardo rojizo y ferralítico rojo; que la vegetación es de mogotes, y se combinan los cultivos varios con pastos y vegetación secundaria.

El depósito paleontológico objeto de estudio es conocido con el nombre de El Abrón. Sin embargo dos depósitos llevan el mismo nombre en el área de estudio, uno es una cueva y el otro es una solapa cársica (véase Martínez, 2003)[2] la cual constituye nuestro objeto de estudio. Según esta fuente esta solapa se encuentra a 250 m s.n.m, en la serie de mapas 1:50 000, hoja Herradura 3583-IV, del Instituto Cubano de Geodesia y Cartografía, 1982 (Suárez y Díaz-Franco, 2003).

### 2.2. Información cartográfica y procesamiento digital de imágenes

El procesamiento de la información cartográfica fue posible mediante la utilización de Sistemas de Información Geográfica (SIG). La imagen satelital de la Sierra de la Güira fue capturada y georreferenciada con el programa SAS Planet Release, versión 150915; la salida cartográfica que se representa en la imagen de localización (Fig. 2) se realizó mediante el programa SAGA, versión 2.2.1. Finalmente, el procesamiento digital del resto de las imágenes que se presentan en el documento se realizó con el software Abobe Photoshop CS5 (versión portable para MS Windows).

### 2.3. Material examinado

Todo el material extraído del depósito fosilífero se encuentra actualmente depositado en el almacén de colecciones paleontológicas número cuatro del Museo Nacional de Historia Natural, La Habana, Cuba [MNHNCu]. De los segmentos estratigráficos identificados el No. VII ha sido el único estudiado exhaustivamente, en el cual se realizaron fechados radiocarbónicos (Suárez y Díaz-Franco, 2003), ubicando cronológicamente el depósito dentro del período Cuaternario (época Pleistoceno tardío) con una antigüedad de 17 406 ±161 años AP. El material paleontológico objeto de estudio del presente trabajo procede específicamente del segmento estratigráfico número nueve (Nivel IX). Las observaciones (tafonómicas) relacionadas con el estado de conservación de los elementos óseos se basan, esencialmente, en los criterios de Holz y Barberena (1994), y Fernández-López (2000a).

[2] Martínez Zamora, L. 2003. Historia Local de Los Palacios. Capítulo II: Período Prehispánico. *En* Mil Cumbres, Folio 3583-IV X470172.

## 2.4. Metodología utilizada para la identificación del material óseo

Limpieza del material

Los fósiles pertenecientes al nivel objeto de estudio se encontraban sin limpiar e identificar. Los mismos estaban almacenados en sacos, los cuales presentaban un número en la superficie que indicaba el nivel estratigráfico del que procedía. Estos materiales estaban mezclados con numerosas partículas sedimentarias de carácter mineral, esencialmente fragmentos de rocas (Fig. 3A), así como una gran cantidad de restos del sedimento en el que se acumularon.

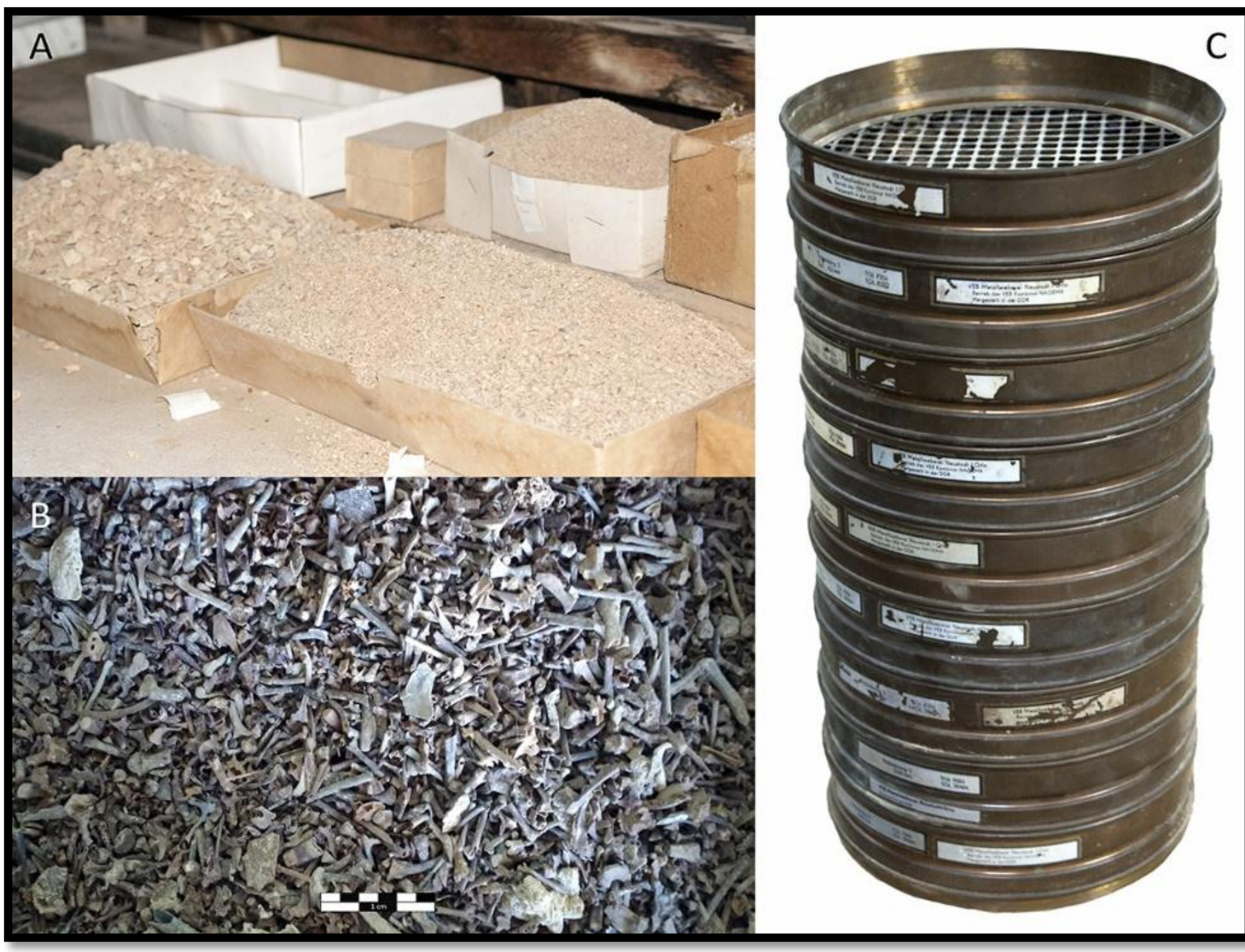

**FIGURA 3**. (**A**) Material residual posterior a la extracción de los materiales óseos. Nótese que se pueden distinguir al menos tres tamaños diferentes de partículas sedimentarias; (**B**) Muestra de algunos materiales óseos recuperados después del proceso de limpieza. Nótese el alto grado de fragmentación y el pequeño tamaño de los restos; cada segmento de la barra inferior de la escala representa 1 cm.; (**C**) Engranaje de los tamices de diferente milimetrado para el proceso de cernido y selección de los materiales óseos.

Esta característica dificultó el trabajo de limpieza debido a que la gran mayoría de los restos óseos extraídos presentan un pequeño tamaño, por proceder mayormente de grupos zoológicos como anfibios, reptiles, y aves, los cuales no son objetivo del presente trabajo, pero formaron parte del material seleccionado, al igual que los restos de mamíferos.

Como casi la totalidad del material óseo se encuentra fragmentado (Fig. 3B) el procedimiento de laboratorio utilizado consistió en el acoplamiento de 9 tamices de diferentes milimetrajes de enrejillado (8,0 mm, 6,3 mm, 5,0 mm, 3,15 mm, 2,5 mm, 1,25 mm, 0,80 mm, 0,40 mm y 0,25 mm), dispuestos de forma ascendente, los más finos se colocaron en la base y los más gruesos en el ápice (Fig. 3C).

Para cernir el material se hicieron vibrar los tamices al unísono de forma tal que los componentes más gruesos se quedaban en la superficie del tamiz y el resto se iba depositando hacia la base según los diferentes grados de enrejillado. Este proceso de filtrado permitió una mejor selección del material. Posterior al proceso de filtrado, el material contenido en cada tamiz era vertido en una bandeja y se procedía a la selección del mismo utilizando pinzas finas, ya que era necesario separar los restos óseos del sedimento. Los materiales de mayor tamaño se colocaron en un recipiente y los de menor tamaño en otro.

Para la selección del material óseo se tuvo en cuenta toda aquella pieza ósea que preservara caracteres morfológicos y parámetros osteométricos que permitieran la identificación de los diferentes taxones. Los desechos de cada tamiz se ubicaron en cajas de cartón agrupadas según el tamaño del desecho. Preservamos los desechos para posibles revisiones por parte de investigadores con mayor experiencia. Luego de seleccionado el material se procedió al lavado del mismo, los elementos óseos fueron depositados en el tamiz con menor grado de enrejillado (0,25 mm) y fueron lavados con agua corriente con un total de diez repeticiones. Después de lavado el material se colocó esparcido en dos bandejas y se dejó secar por 5 días a temperatura ambiente, cuando estuvo seco se colocaron en dos recipientes diferentes de acuerdo con el tamaño de los restos. Posteriormente los huesos fueron separados por tipo, especies y lateralidad.

Criterios taxonómicos

Para realizar la identificación de las especies de mamíferos presentes en el depósito se utilizaron criterios de varios autores, atendiendo a la composición zoológica hallada en el nivel objeto de estudio.

El trabajo de Silva (1979) fue la base esencial para la identificación de especies del orden Chiroptera (murciélagos), aunque la especie *Phyllops silvai*, en particular, requirió de la consulta del trabajo de Suárez y Díaz-Franco (2003) donde fue descrita la especie. De ambas fuentes se tomaron todos los *caracteres morfológicos* y *parámetros osteométricos* necesarios para la labor taxonómica en este orden. Los cambios sistemáticos acontecidos en los últimos años para este grupo se consultaron del trabajo de Silva y Vela (2010) [3].

Para el caso de los órdenes Soricomorpha (musarañas), y Rodentia (jutías y ratas espinosas) fue utilizado en mayor medida el criterio propuesto en Silva *et al.* [2008], obra ésta, hasta el momento, la de mayor actualización taxonómica para la fauna de mamíferos extintos del archipiélago cubano, solo con la excepción de los mamíferos voladores. Para todos los casos se trató taxonómicamente el estudio hasta el nivel de género, y en los casos posibles hasta el nivel de especie.

Como una generalidad para el proceso de identificación, constituye un problema en los tres órdenes presentes la cantidad de piezas óseas con alto grado de fragmentación. Esto trajo como consecuencia la imposibilidad de establecer comparaciones morfológicas en algunos casos con las fuentes antes mencionadas, además de que, paralelamente, se suma la ausencia de los parámetros osteométricos, necesarios para la identificación hasta el nivel de especie.

En el sentido de aminorar este problema acudimos también a criterios de otros especialistas, tales como el Dr. Carlos Arredondo Antúnez, de la Facultad de Biología, y Lázaro W. Viñola López, de la Facultad de Geografía, ambos de la Universidad de La Habana, también de Osvaldo

[3] SILVA, G.; Y H. VELA. 2010. Actualización taxonómica y distribucional de los murciélagos de Cuba. En Biodiversidad Paleontológica del Archipiélago Cubano: Bases Cartográficas y Conservacionistas. Informe final del proyecto 022 AMA-CITMA; 074 MNHN. (CD ROM, inédito). 9pp.

Jiménez Vásquez, de la Oficina del Historiador de La Habana, los cuales ayudaron en la identificación de especies.

Para todos los órdenes estudiados las piezas óseas que se tuvieron en cuenta para la identificación de género y especie fueron: cráneo, hemimandíbula, húmero, y fémur; con la excepción del orden Chiroptera, en el cual solo se consideró para su estudio las piezas hemimandibulares. Tanto los *caracteres morfológicos* como los *parámetros osteométricos* tenidos en cuenta para la identificación de las especies fueron tomados de las referencias antes mencionadas, las cuales no presentaban abreviaturas originalmente. Por lo tanto, en el presente trabajo se decidió crear las abreviaturas para estos caracteres y parámetros en función de facilitar la comprensión de los resultados expuestos, fundamentalmente en las tablas, así como su discusión.

Para el caso particular de las tablas se presentó el problema de la disposición de los datos, los cuales no fue posible presentar el mismo tipo, uniformemente, en sentido horizontal o vertical, debido a que sobrepasaban el espacio adecuado de la caja tipográfica del documento. Es decir, se tuvo que alternar la relación carácter morfológico o parámetro osteométrico con respecto a las piezas óseas, para evitar el problema mencionado anteriormente. Como este aspecto puede atentar contra la ergonomía del documento, en aras de aminorarlo se decidió asignar colores a las variables carácter morfológico [Azul, R(222) G(234), B(246)] y parámetros osteométricos [Verde, R(197) G(274), B(179)] para facilitar la relación visual entre cada carácter o parámetro con el valor de la pieza en cada tabla.

En el caso de las figuras 4, y 5, fueron tomadas y modificadas de Silva *et al.* [2008]; en ellas se reflejan los *caracteres morfológicos* (en números) y *parámetros osteométricos* (en letras) señalados con saetas y líneas discontinuas respectivamente.

Todas las medidas se tomaron con un pie de rey, error 0,02 mm. Se analizaron 22 caracteres morfológicos para cráneo, hemimandíbula, húmero y fémur, siguiendo el criterio del autor mencionado anteriormente.

**ORDEN SORICOMORPHA**

En esta especie se tomaron 6 mediciones para cráneo, 36 para hemimandíbula, 75 para húmero y 90 para fémur arrojando un total de 207 mediciones. Todos los parámetros osteométricos tomados fueron propuestos por Silva *et al.* [2008], (Fig. 4). Para el caso del cráneo no se tuvo en cuenta los caracteres morfológicos ni parámetros osteométricos debido a la total ausencia de la región posterior en los mismos. Por ende la Fig.4 (**A, B, C**) que muestra las vistas para esta pieza no incluye esta región.

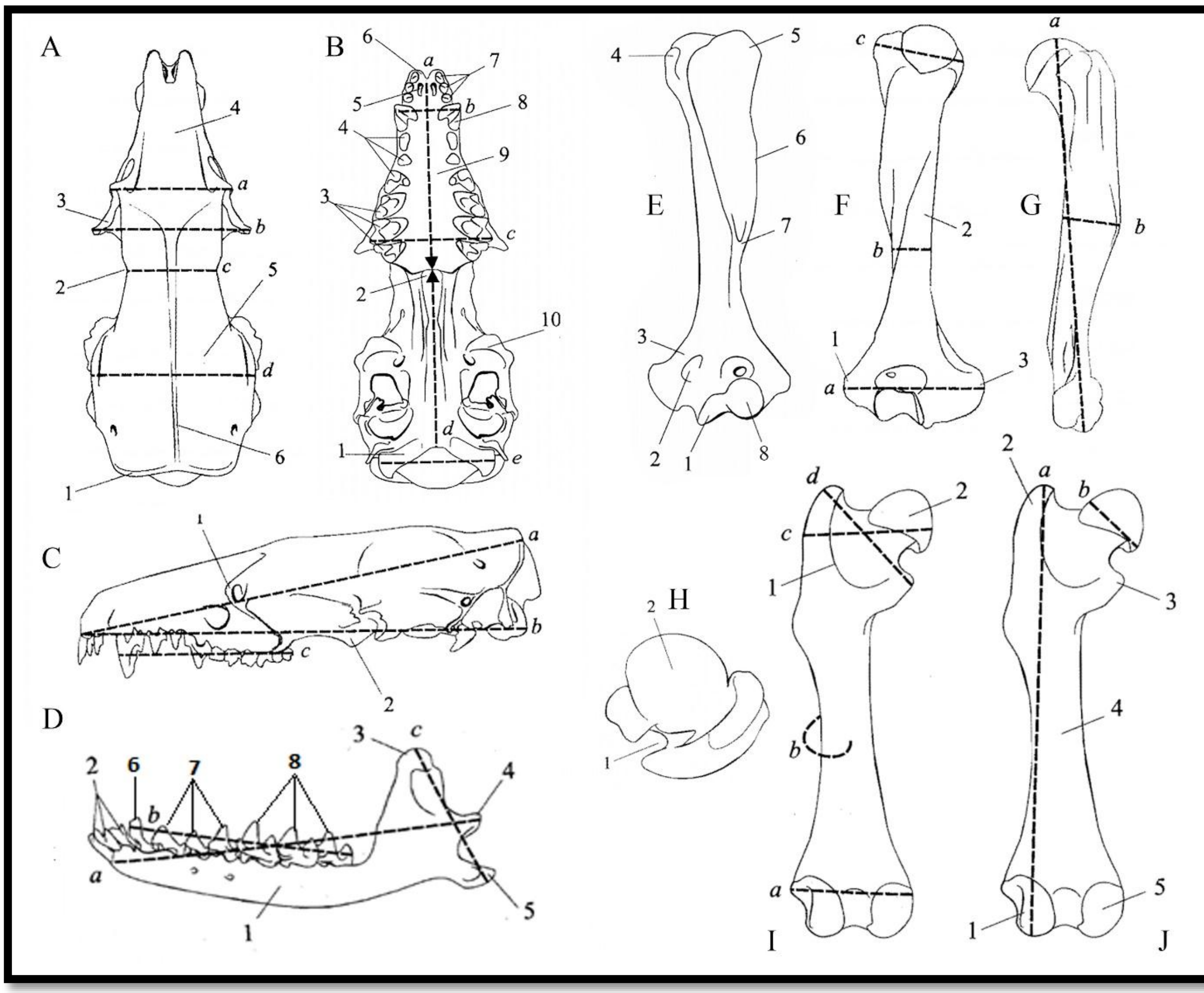

**FIGURA 4.** *Caracteres morfológicos* y *parámetros osteométricos* para el Orden Soricomorpha (Sp. *Nesophontes micrus*, específicamente). (**A-C**) Vistas del Cráneo: superior (**A**), inferior (**B**), y lateral (**C**); (**D**) Hemimandíbula en vista lateral (cara labial); (**E-H**) Vistas del Húmero: anterior (**E**), posterior (**F**), lateral interna (**G**), superior (**H**); (**I-J**) Vista posterior del Fémur. Tomado y modificado de Silva *et al*. [2008].

## Cráneo (vistas)

Fig. 4(A). *Caracteres morfológicos*: (1) Cresta lamboidea [**CLmb**]; (2) Constricción postorbital [**CPorb**]; (3) Proceso cigomático del maxilar [**PrCigMax**]; (4) Rostro [**Rs**]; (5) Bóveda craneana [**BC**]; (6) Cresta sagital [**CrSag**]. *Parámetros osteométricos*: (a) Anchura lacrimal: Distancia entre los puntos más láteroexternos de los lacrimales [**AnLcr**]; (c) Anchura postorbital: Distancia (transversal) mínima entre puntos opuestos de la constricción postorbital [**AnPorb**].

Fig. 4(B). *Caracteres morfológicos*: (1) Cóndilo occipital [**ConOcc**]; (2) Fosa mesopterigoidea [**FsMpt**]; (3) Molares [$\mathbf{M^1, M^2, M^3}$]; (4) Premolares [$\mathbf{P^{1,} P^2, P^3}$]; (5) Foramen incisivo [**ForInc**]; (6) Premaxilar [**Pmx**]; (7) Incisivos [**Inc**]; (8) Canino [**Can**]; (9) Paladar [**Pl**]; (10) Fosa glenoidea [**FsGlen**]. *Parámetros osteométricos*: (a) Longitud palatina: Distancia (axial) entre los bordes anterior y posterior del paladar [**LPl**]; (b) Anchura canina coronaria: Distancia entre los puntos más látero externos de los caninos [**ACanCor**]. Anchura canina alveolar: Igual que la anchura canina coronaria, pero considerando los alveolos (en lugar de las coronas) de ambos dientes [**ACanAlv**]; (c) Anchura molar coronaria: Distancia entre los puntos más láteroexternos de los molares [**AMCor**]. Anchura molar alveolar: Igual que la anchura molar coronaria, pero considerando los alveolos (en lugar de las coronas) de ambos dientes [**AMAlv**].

Fig. 4(**C**). Caracteres morfológicos: (1) Lacrimal [**Lcr**]; (2) Proceso pterigiodeo [**Prtg**].

## Hemimandíbula

Fig. 4(D). *Caracteres morfológicos*: (1) Rama Horizontal [**RHz**]; (2) Incisivos [**Inc**]; (3) Proceso coronoides [**PrCor**]; (4) Cóndilo [**Con**]; (5) Proceso angular [**PrAn**]; (6) Canino [**Can**]; (7) Premolares [$\mathbf{P^{1,} P^2, P^3}$]; (8) Molares [$\mathbf{M^1, M^2, M^3}$]. *Parámetros osteométricos*: (a) Longitud condilar: Distancia entre el punto más anterior del alveolo del primer incisivo y el punto más posterior del cóndilo [**LCon**]; (b) Longitud caninomolar coronaria: Distancia entre el punto más anterior del canino y el punto más posterior del último molar [**LCanMCor**], o Longitud caninomolar alveolar: Igual que la longitud caninomolar coronaria, pero considerando los alveolos (en lugar de las coronas) de ambos dientes [**LCanMAlv**]; (c) Altura ángulocoronoidea: Distancia máxima entre el proceso angular y el extremo superior del proceso coronoides [**AlCor**].

**Húmero (vistas)**

Fig. 4(**E**). *Caracteres morfológicos*: (1) Tróclea [**Tr**]; (2) Foramen entepicondilar [**FEcon**]; (3) Puente entepicondilar [**PEcon**]; (4) Tubérculo menor [**TuMe**]; (5) Tubérculo mayor [**TuMy**]; (6) Cresta deltoidea [**CDel**]; (7) Eminencia deltopectoral [**EDp**]; (8) Capítulo [**Cap**].

Fig. 4(**F**). *Caracteres morfológicos*: (1) Epicóndilo externo (lateral) [**EconEx**]; (2) Diáfisis [**Df**]; (3) Epicóndilo interno (medial) [**EconInt**]. *Parámetros osteométricos*: (a) Anchura distal: Distancia entre los puntos más laterales de los epicóndilos [**AnDis**]; (b) Anchura de la diáfisis: Distancia mínima entre los bordes laterales de la diáfisis en su zona media [**AnDf**]; (c) Anchura proximal: Distancia entre los puntos más laterales de los tubérculos [**AnProx**].

Fig. 4(**G**). *Parámetros osteométricos*: (a) Longitud total: Distancia entre el punto más proximal y el punto más distal del hueso [**LT**]; (b) Profundidad deltoidea: Distancia (transversal) entre el punto más ánterodistal de la cresta deltoidea y el borde posterior de la diáfisis [**PDel**].

Fig. 4(**H**). *Caracteres morfológicos*: (1) Surco intertubercular (**SInt**); (2) Cabeza (**Cab**).

**Fémur (vistas)**

Fig. 4(**I**). *Caracteres morfológicos* en vista posterior: (1) Cresta intertrocantérica [**CIntr**]; (2) Cabeza [**Cab**]. *Parámetros osteométricos*: (a) Anchura distal: Distancia (transversal) entre los puntos más laterales del extremo distal del hueso [**AnDis**]; (b) Profundidad de la diáfisis: Distancia entre las caras anterior y posterior de la diáfisis, en la zona media del hueso [**PfDf**]; (c) Anchura proximal: Distancia máxima entre los puntos más laterales de la cabeza y del trocánter mayor [**AnProx**]; (d) Longitud intertrocantérica: Distancia (diagonal) máxima entre el trocánter mayor y el trocánter menor [**LInter**].

Fig. 4(**J**). *Caracteres morfológicos* en vista posterior: (1) Cóndilo externo (lateral) [**ConExt**]; (2) Trocánter mayor [**TrMy**]; (3) Trocánter menor [**TrMe**]; (4) Diáfisis [**Df**]; (5) Cóndilo interno (medial) [**ConInt**]. *Parámetros osteométricos*: (a) Longitud total: Distancia máxima entre el trocánter mayor y el cóndilo del mismo lado [**LT**]; (b) Diámetro mayor de la cabeza: Distancia máxima entre puntos opuestos de la circunferencia de la cabeza [**DMyC**].

**ORDEN RODENTIA**

En las piezas estudiadas para este orden se tomaron un total de 117 mediciones (58 para hemimandíbula, 34 para húmero y 25 para fémur), todas propuestas en Silva *et al.* [2008], (Fig. 5).

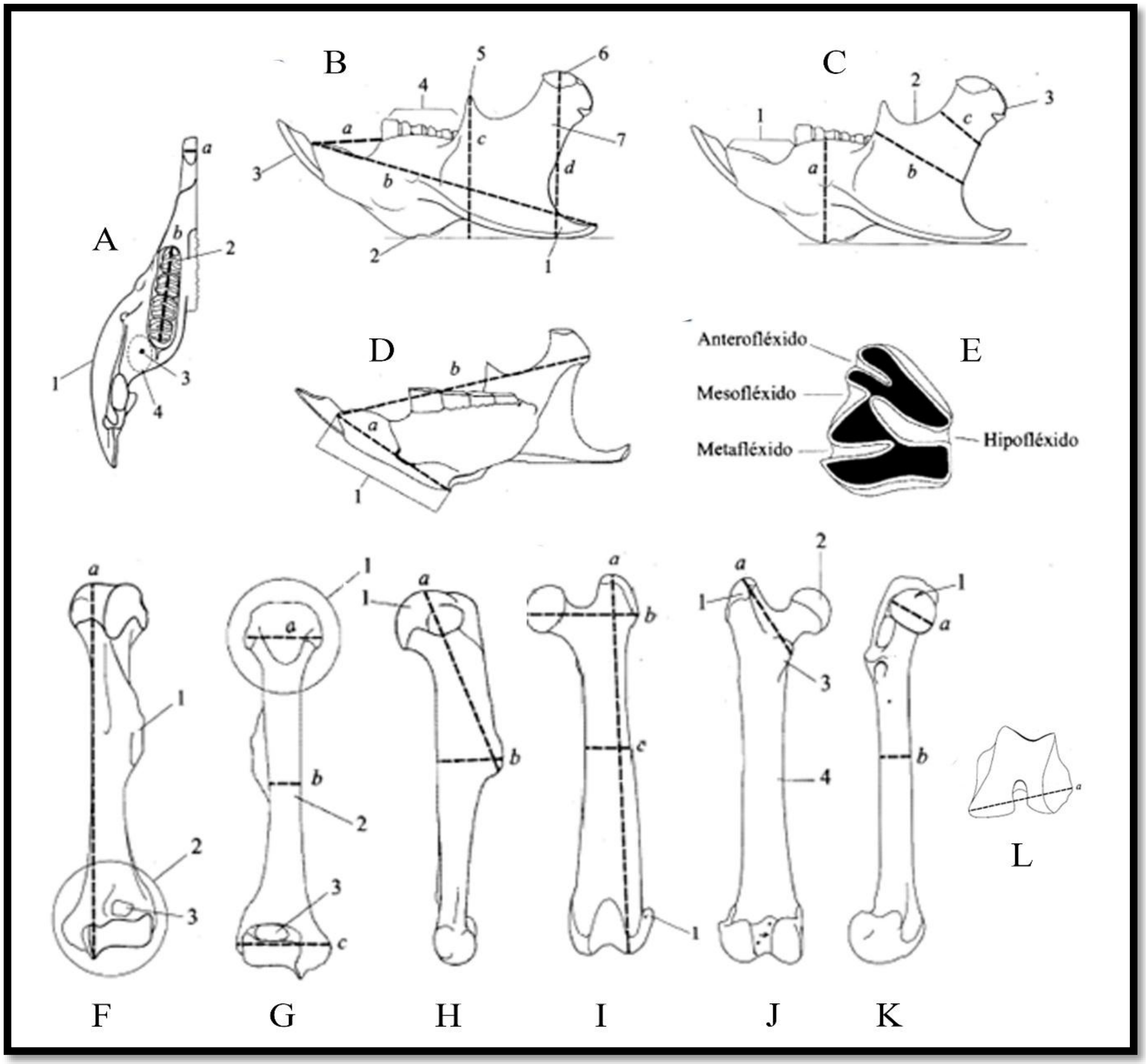

**FIGURA 5**. Caracteres morfológicos y parámetros osteométricos para el Orden Rodentia (jutías y ratas espinosas). (**A-E**) Vistas de la Hemimandíbula: superior (**A**), lateral (cara labial) (**B y C**), lateral (cara lingual) (**D**); Vista del premolar: superior (**E**); (**F-H**) Vistas del Húmero: anterior (**F**), posterior (**G**), lateral (**H**); (**I-K**) Vistas del Fémur: anterior (**I**), posterior (**J**), lateral (**K**), inferior (**L**). Tomado y modificado de Silva *et al*. [2008].

**Hemimandíbula (vistas)**

Fig. 5 (**A**). *Caracteres morfológicos* en vista superior: (1) Cresta masetérica [**CrMas**]; (2) Premolar [**P**]; (3) Foramen mandibular [**ForMb**]; (4)  Fosa retromolar [**ForRet**]. *Parámetros*

*osteométricos*: (a) Anchura coronaria incisiva: Distancia entre los puntos más laterales del incisivo [**AnCorInc**]; (b) Longitud coronaria de la serie molariforme: Distancia máxima entre las coronas de los dientes primero y último de la serie [**LCorSMol**]. Longitud alveolar de la serie molariforme: Distancia máxima entre las coronas de los dientes primero y último de la serie [**LAlvSMol**].

Fig. 5 (**B**). *Caracteres morfológicos* en vista lateral (cara labial): (1) Proceso angular [**PrAn**]; (2) Proceso mental [**PrMen**]; (3) Incisivo [**Inc**]; (4) Molariformes [**M**]; (5) Proceso coronoides [**PrCor**]; (6) Cóndilo [**Con**]; (7) Proceso condiloides [**PrCon**]. *Parámetros osteométricos*: (a) Longitud de la diastema: Distancia entre el punto más posterior del borde alveolar del incisivo y el punto más anterior del borde alveolar del premolar [**LDia**]; (b) Longitud ángulosinficial: Distancia entre el punto más posterior del proceso angular y el punto más anterosuperior de la sínfisis [**LAns**]; (c) Altura coronoidea: Distancia (perpendicular) entre el punto más alto del proceso coronoides y la línea imaginaria que une los puntos más inferiores del proceso mental y el proceso angular [**AlCor**]; (d) Altura condilar: Distancia (perpendicular) entre el punto más alto del cóndilo y la línea imaginaria que une los puntos más inferiores del proceso mental y el proceso angular [**AlCon**].

Fig. 5 (**C**). *Caracteres morfológicos* (adicionales) en vista lateral (cara labial): (1) Diastema [**Dst**]; (2) Escotadura sigmoidea [**EsSg**]; (3) Proceso postcondiloides [**PrPstcon**]. *Parámetros osteométricos*: (a) Altura corporal: Distancia entre el punto más inferior del proceso mental y el punto del borde alveolar labial que marca la separación entre el premolar y el primer molar [**AlCorp**]; (b) Anchura mayor de la rama ascendente: Distancia mínima entre el borde anterior del proceso coronoides y el borde posterior de la rama ascendente [**AnMyRmAs**]; (c) Anchura menor de la rama ascendente: Distancia mínima entre la escotadura sigmoidea y el borde posterior de la rama ascendente [**AnMeRmAs**].

Fig. 5 (**D**). *Caracteres morfológicos* en vista lateral (cara lingual): (1) Sínfisis [**Sf**]. *Parámetros osteométricos*: (a) Longitud de la sínfisis: Distancia (axial) entre los extremos anterosuperior y pósteroinferior de la sínfisis [**LSf**]; (b) Longitud cóndilosinficial: Distancia entre el punto más posterior del proceso postcondiloides y el punto más anterosuperior de la sínfisis [**LCons**].

Fig. 5 (**E**). *Caracteres morfológicos* generales del premolar.

## Húmero (vistas)

Fig. 5 (**F**). *Caracteres morfológicos* en vista anterior: (1) Cresta deltoidea [**CrDel**]; (2) Extremo distal [**ExDstl**]; (3) Fosa coronoidea [**FsCor**]. *Parámetros osteométricos*: (a) Longitud total: Distancia entre el punto más proximal y el punto más distal del hueso [**LT**].

Fig. 5 (**G**). *Caracteres morfológicos* en vista posterior: (1) Extremo proximal [**ExProx**]; (2) Diáfisis [**Df**]; (3) Fosa olecraneana [**FsOlcr**]. *Parámetros osteométricos*: (a) Anchura proximal: Distancia entre los puntos más laterales de los tubérculos [**AnProx**]; (b) Anchura de la diáfisis: Distancia mínima entre los bordes laterales de la diáfisis en su zona media [**AnDf**]; (c) Anchura distal: Distancia entre los puntos más laterales de los epicóndilos [**ADis**].

Fig. 5(**H**). *Caracteres morfológicos* en vista lateral interna: (1) Cabeza [**Cab**]. *Parámetros osteométricos*: (a) Longitud deltoidea: Distancia máxima entre la cabeza y el punto más ánteroinferior de la cresta deltoidea [**LDel**] (b) Profundidad deltoidea: Distancia (transversal) entre el punto más ánterodistal de la cresta deltoidea y el borde posterior de la diáfisis [**PfDel**].

## Fémur (vistas)

Fig. 5(**I**). *Caracteres morfológicos* en vista anterior: (1) Cóndilo externo [**ConExt**]. *Parámetros osteométricos*: (a) Longitud total: Distancia máxima entre el trocánter mayor y el cóndilo del mismo lado [**LT**]; (b) Anchura proximal: Distancia máxima entre los puntos más laterales de la cabeza y del trocánter mayor [**AnProx**]; (c) Anchura diáfisis: Distancia mínima (transversal) entre los bordes laterales de la diáfisis en el punto medio [**AnDf**].

Fig. 5(**J**). *Caracteres morfológicos* en vista posterior: (1) Trocánter mayor [**TrMy**]; (2) Cabeza [**Cab**]; (3) Trocánter menor [**TrMn**]; (4) Diáfisis [**Df**]. *Parámetros osteométricos*: (a) Longitud intertrocantérica: Distancia (diagonal) máxima entre el trocánter mayor y el trocánter menor [**LIntr**].

Fig. 5(**K**). *Caracteres morfológicos* en vista anterior: (1) Fosita para el ligamento redondo (fovea) [**FsLigR**]. *Parámetros osteométricos*: (a) Diámetro mayor de la cabeza: Distancia máxima entre puntos opuestos de la circunferencia de la cabeza [**DMyCab**]; (b) Profundidad

de la diáfisis: Distancia entre las caras anterior y posterior de la diáfisis, en la zona media del hueso [**PfDf**].

Fig. 5(**L**). *Parámetros osteométricos* en vista inferior: (a) Anchura distal: Distancia (transversal) entre los puntos más laterales del extremo distal del hueso (a nivel de los cóndilos).

**ORDEN CHIROPTERA**

Los murciélagos fueron identificados según los criterios propuestos por Silva (1979), y Suárez y Díaz-Franco (2003), utilizando hemimandíbulas. La variable medida fue la longitud mentoniana (**LM**) en aquellas hemimandíbulas que preservaban los caracteres morfológicos que permitían la toma de esta medida, (Fig.6).

**Hemimandíbula**

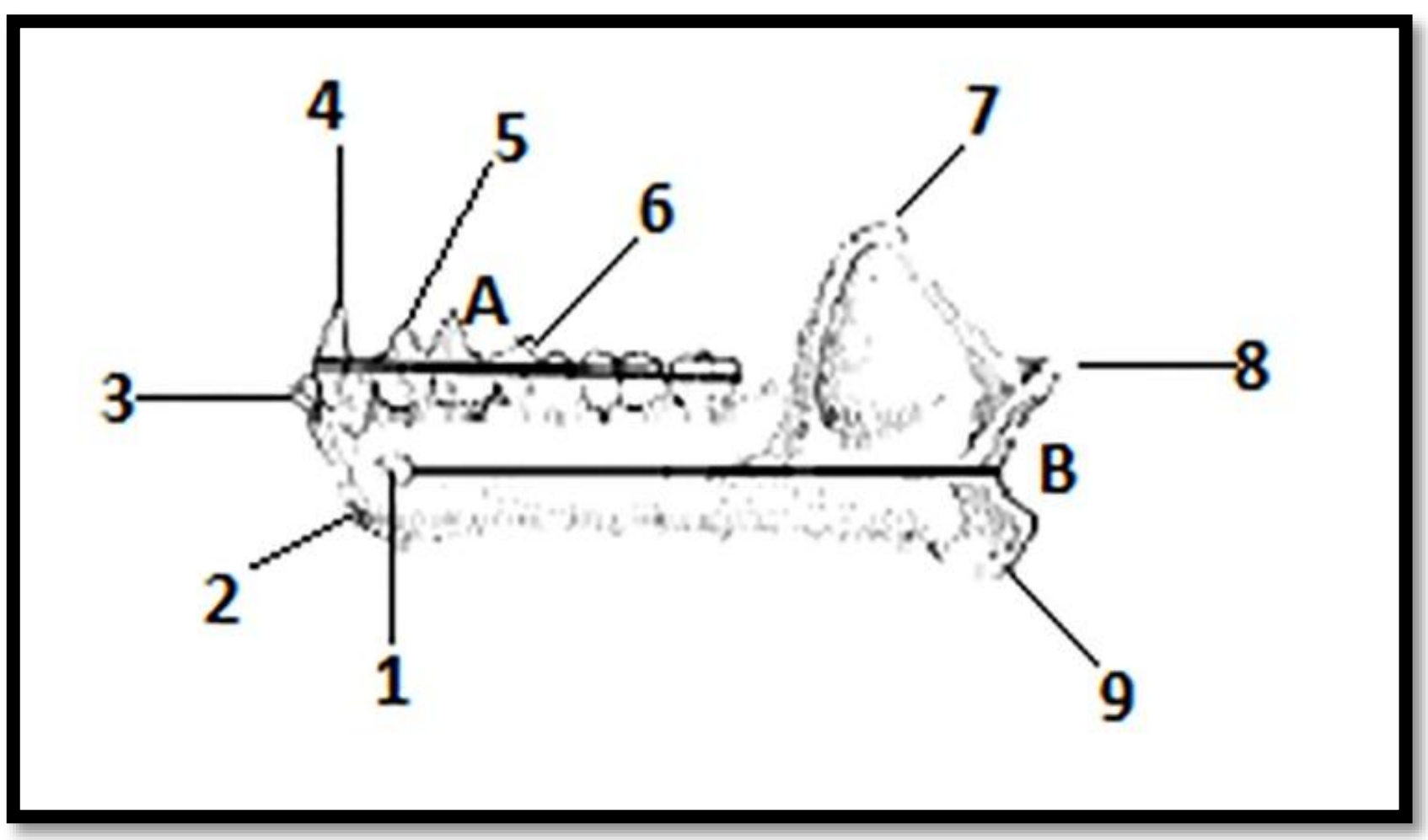

**FIGURA 6.** Caracteres morfológicos y parámetros osteométricos para el Orden Chiroptera sólo para hemimandíbulas en vista lateral externa (cara labial). Tomado y modificado de Silva (1979).

Fig. 6. *Caracteres morfológicos* en vista lateral externa (cara labial): (1) Foramen mentoniano [**ForMen**]; (2) Sínfisis manbdibular [**SfMb**]; (3) Incisivos [**Inc**]; (4) Canino [**Can**]; (5) Premolares [**$P^1$, $P^2$**]; (6) Molares [**$M^1$, $M^2$, $M^3$**]; (7) *Proceso coronoides* [**PrCor**]; (8) Cóndilo mandibular [**ConMb**]; (9) Proceso angular [**PrAn**]. *Parámetros osteométricos*: (**A**) Longitud canino molar [**LCanM**]: Distancia, en el eje de la hilera dental, entre el punto más anterior del canino y el punto más posterior del último molar; (b) Longitud mentoniana [**LMen**]: (**B**) Distancia mínima entre el foramen mentoniano y la escotadura que media entre el cóndilo mandibular y el proceso angular.

# CAPÍTULO 3. RESULTADOS

## 3.1. Taxonomía del depósito

La fauna de mamíferos hallados en el nivel objeto de estudio (Tabla I) del depósito paleontológico El Abrón, Los Palacios, Pinar del Río, Cuba, está compuesta esencialmente por 3 órdenes, 7 familias y 14 especies, dentro de los cuales algunos de los restos pertenecen a especies que aún viven en nuestro archipiélago [representadas con un asterisco (*)]. El orden más representativo es Chiroptera (fauna de murciélagos), el cual presenta 4 familias, 9 géneros y 9 especies del total identificadas.

También en este resultado se hace alusión a especies de mamíferos que no habían sido identificadas en estudios anteriores (véase Suárez y Díaz-Franco, 2003; Silva *et al.* [2008]; Suárez y Díaz-Franco, 2011), las cuales representaremos con la letra **(n)** después del nombre de la especie.

**TABLA I**. Material osteológico analizado para la determinación de la composición taxonómica de la fauna de mamíferos hallados en el depósito.TP=Tipo de pieza; CPOE=Cantidad de piezas óseas estudiadas; D=Derecho; I=Izquierdo; RC=Restos completos; RF=Restos fragmentados; NMI=Número mínimo de individuos.

| TP | CPOE | Lateralidad | | Conservación | | NMI |
|---|---|---|---|---|---|---|
| | | D | I | RC | RF | |
| | | | ***Nesophontes micrus*** | | | |
| Cráneos | 3 | - | - | - | 3 | |
| Hemimandíbulas | 138 | 71 | 67 | 25 | 113 | |
| Húmeros | 184 | 111 | 73 | 117 | 67 | 111 |
| Fémures | 176 | 84 | 92 | 74 | 102 | |
| | | | ***Boromys offella*** | | | |
| Hemimandíbulas | 2 | - | 2 | - | 2 | |
| Húmeros | 8 | 5 | 3 | 3 | 5 | |
| Fémures | 16 | 12 | 4 | 8 | 8 | 12 |
| | | | ***Boromys torrei*** | | | |
| Hemimandíbulas | 33 | 14 | 19 | 2 | 31 | |
| Húmeros | 37 | 16 | 21 | 24 | 13 | 21 |
| Fémures | 15 | 9 | 6 | 9 | 6 | |
| | | | ***Geocapromys columbianus*** | | | |
| Hemimandíbulas | 5 | 3 | 2 | - | 5 | 3 |
| | | | ***Mesocapromys* sp. (n)** | | | |
| Hemimandíbulas | 1 | - | 1 | - | 1 | 1 |
| | | | ***Brachyphylla nana**** | | | |
| Hemimandíbulas | 2 | - | 2 | - | 2 | 2 |
| | | | ***Erophylla sezekorni** (n)** | | | |
| Hemimandíbulas | 2 | - | 2 | - | 2 | 2 |
| ***Phyllonycteris poeyi**** | | | | | | |
| Hemimandíbulas | 2 | 1 | 1 | 1 | 1 | 2 |
| ***Macrotus waterhousei**** | | | | | | |
| Hemimandíbulas | 29 | 15 | 14 | 10 | 19 | 15 |
| ***Monophyllus redmani*** **(n)** | | | | | | |
| Hemimandíbulas | 1 | - | 1 | 1 | - | 1 |
| ***Eptesicus fuscus**** | | | | | | |
| Hemimandíbulas | 2 | - | 2 | 2 | - | 2 |
| ***Pteronotus parnelli*** **(n)** | | | | | | |
| Hemimandíbulas | 2 | 2 | - | 1 | 1 | 2 |
| ***Tadarida brasiliensis*** **(n)** | | | | | | |
| Hemimandíbulas | 2 | 2 | - | 1 | 1 | 2 |
| ***Phyllops silvai*** | | | | | | |
| Hemimandíbulas | 1 | - | 1 | 1 | - | 1 |

CLASIFICACIÓN POR GRUPOS ZOOLÓGICOS (* Especies vivientes)

Grupo Zoológico #1 (GZ#1)
**Orden S O R I C O M O R P H A**
Superfamilia SORICOIDEA
Familia NESOPHONTIDAE
Género *Nesophontes*
*Nesophontes micrus*

Grupo Zoológico #2 (GZ#2)
**Orden R O D E N T I A**
Suborden HYSTRICOGNATHA
Infraorden HYSTRICOGNATHI
Parvorden CAVIIDA
Superfamilia CAVIOIDEA
Familia ECHIMYIDAE
Subfamilia HETEROPSOMYINAE
Género *Boromys*
*Boromys offella*
*Boromys torrei*
Familia CAPROMYIDAE
Subfamilia CAPROMYINAE
Género *Geocapromys*
*Geocapromys columbianus*
Género *Mesocapromys*
*Mesocapromys* sp.

Grupo Zoológico #3 (GZ#3)
**Orden C H I R 0 P T E R A**
Suborden MICROCHIROPTERA
Superfamilia PHYLLOSTOMATOIDEA

Familia MORMOOPIDAE
Género *Pteronotus*
Subgénero *Phyllodia*
*Pteronotus parnelli**
Familia PHYLLOSTOMIDAE
Subfamilia PHYLLOSTOMINAE
Género *Macrotus*
*Macrotus waterhousei**
Subfamilia BRACHYPHYLLINAE
Género *Brachyphylla*
*Brachyphylla nana**
Género *Erophylla*
*Erophylla sezekorni sezekorni**
Género *Phyllonycteris*
Subgénero *Phyllonycteris*
*Phyllonycteris poeyi* *
Subfamilia GLOSSOPHAGINAE
Género *Monophyllus*
*Monophyllus redmani**
Subfamilia STENODERMATINAE
Género *Phyllops*
*Phyllops silvai*
Familia VESPERTILIONIDAE
Subfamilia VESPERTILIONINAE
Género *Eptesicus*
*Eptesicus fuscus**
Familia MOLOSSIDAE
Subfamilia MOLOSSINAE
Género *Tadarida*
*Tadarida brasiliensis**

**3.2. Estado de conservación de los restos óseos**

Constituye una generalidad que tanto los caracteres morfológicos así como los parámetros osteométricos conservados en las piezas óseas correspondientes a los órdenes Soricomorpha (GZ#1) y Rodentia (GZ#2), permitieron la determinación del género (para todos los casos) y las especies (con excepción del género *Mesocapromys*), siguiendo el criterio propuesto en Silva *et al.* [2008], y el criterio de Arredondo (2015, *com. pers*). Por otra parte, los caracteres morfológicos y los parámetros osteométricos conservados en las piezas óseas correspondientes al orden Chiroptera (GZ#3) permitieron la determinación del género y especies para todos los casos siguiendo los criterios propuestos en Silva (1979), y Suárez y Díaz-Franco (2003), además de las actualizaciones sistemáticas consideradas para este grupo en el trabajo de Silva y Vela (2010).

GZ#1: ***Nesophontes micrus***

**Cráneos**

**TABLA II**. Caracteres morfológicos conservados (celdas sombreadas) para los cráneos de *N. micrus.* T=Completos; P=Parcial; I=Izquierdo; D=Derecho.

| *CARAC. MORF.* | *NMC_01* | *NMC_02* | *NMC_03* |
|---|---|---|---|
| **CrLmb** | | | |
| **CPorb** | T | | |
| **PrCigMax** | T (I) | T(I) | |
| **Rs** | T | T | T |
| **BC** | P | | |
| **CrSag** | | | |
| **ConOcc** | | | |
| **FsMpt** | T | | |
| **ForInc** | | T (D, I) | T (D, I) |
| **Pmx** | | | |
| **Pl** | | P | T |
| **FsGlen** | | | |
| **Inc** | | | |
| **Can** | | T (D, I) | |
| **$P^1$** | | I | |
| **$P^2$** | I | | |
| **$P^3$** | | | |
| **$M^1$** | | I | |
| **$M^2$** | I | | |
| **$M^3$** | | | |
| **Lcr** | T (D , I) | T (I) | P (D) |
| **Prtg** | | | |

**TABLA III**. Parámetros osteométricos conservados para los cráneos de *N. micrus.*

| *PARÁM. OSTEOM.* | *NMC_01* | *NMC_02* | *NMC_03* |
|---|---|---|---|
| **AnLcr** | 8,2 | - | - |
| **AnPporb** | 6,68 | - | - |
| **LP** | 11,48 | - | - |
| **ACanCor** | - | 4,52 | - |
| **ACanAlv** | 4,2 | - | 3,88 |
| **AMCor** | - | - | - |
| **AMAlv** | - | - | - |

## Hemimandíbulas

**TABLA IV**. Caracteres morfológicos conservados (celdas sombreadas) para las hemimandíbulas de *N. micrus.* T=Completos; P=Parcial.

| *CARAC. MORF.* | RHz | Prcor | Con | Pran | Inc | Can | $P^1$ | $P^2$ | $P^3$ | $M^1$ | $M^2$ | $M^3$ |
|---|---|---|---|---|---|---|---|---|---|---|---|---|
| ***NMHM_01*** | T | T | T | T | | | | | | | | |
| ***NMHM_02*** | T | T | T | T | | | | | | | | |
| ***NMHM_03*** | T | T | T | T | | | | | | | | |
| ***NMHM_04*** | T | T | T | T | | | | | | | | |
| ***NMHM_05*** | P | T | T | T | | | | | | | | |
| ***NMHM_06*** | P | T | T | T | | | | | | | | |
| ***NMHM_07*** | P | T | T | T | | | | | | | | |
| ***NMHM_08*** | P | T | T | T | | | | | | | | |
| ***NMHM_09*** | T | T | T | T | | | | | | | | |
| ***NMHM_10*** | P | T | T | T | | | | | | | | |
| ***NMHM_11*** | T | T | T | T | | | | | | | | |
| ***NMHM_12*** | P | T | T | T | | | | | | | | |
| ***NMHM_13*** | T | T | T | T | | | | | | | | |
| ***NMHM_14*** | P | T | T | T | | | | | | | | |
| ***NMHM_15*** | T | T | T | T | | | | | | | | |

**TABLA V.** Parámetros osteométricos conservados para las hemimandíbulas de *N. micrus*

| *PARÁM. OSTEOM.* | AlCor | LCMAlv | LCMCor | LCon |
|---|---|---|---|---|
| ***NMHM_01*** | 7,58 | 10,82 | - | 18,3 |
| ***NMHM_02*** | 7,26 | 9,92 | - | 17,78 |
| ***NMHM_03*** | 8,24 | 10,8 | - | 17,58 |
| ***NMHM_04*** | 7,86 | 9,32 | - | 16,48 |
| ***NMHM_05*** | 7,1 | 8,7 | - | - |
| ***NMHM_06*** | 7,96 | 8,9 | - | 16,5 |
| ***NMHM_07*** | 9,38 | 11,08 | - | 19,16 |
| ***NMHM_08*** | 7,12 | 8,34 | - | 16,36 |
| ***NMHM_09*** | 8,7 | - | 11,3 | 18,44 |
| ***NMHM_10*** | 7,64 | - | - | - |
| ***NMHM_11*** | 9,08 | 10,82 | - | 19,34 |
| ***NMHM_12*** | 8,52 | - | - | - |
| ***NMHM_13*** | 7,6 | - | - | - |
| ***NMHM_14*** | 8,7 | 10,86 | - | 18,3 |
| ***NMHM_15*** | 9,06 | - | - | - |

## Húmeros

Todos los húmeros (NMHU_01 – NMHU_15) corresponden a ejemplares adultos y son de lateralidad derecha. Presentan un buen estado de conservación, con la presencia de los caracteres morfológicos por región del hueso, aptos para el análisis osteométrico. La región de la epífisis proximal conserva: surco intertubercular (**SInt**), cabeza (**Cab**), tubérculo menor (**TuMe**), tubérculo mayor (**TuMy**), y cresta deltoidea (**CDel**); la región distal conserva: epicóndilo externo (**EconEx**), epicóndilo interno (**EconInt**), tróclea (**Tr**), foramen entepicondilar (**FEcon**), puente entepicondilar (**PEcon**) y capítulo (**Cap**).

**TABLA VI.** Parámetros osteométricos conservados para los húmeros de *N. micrus*

| ***PARÁM. OSTEOM.*** | **AnDf** | **AnDis** | **AnProx** | **LT** | **PDel** |
|---|---|---|---|---|---|
| ***NMHU_01*** | 1,26 | 4,9 | 3,48 | 16,06 | 2,06 |
| ***NMHU_02*** | 1,18 | 3,7 | 2,88 | 14,02 | 1,8 |
| ***NMHU_03*** | 1,18 | 3,7 | 2,86 | 14,02 | 1,8 |
| ***NMHU_04*** | 1,56 | 4,12 | 3,58 | 16,2 | 2,04 |
| ***NMHU_05*** | 1,2 | 3,64 | 2,76 | 13,12 | 1,68 |
| ***NMHU_06*** | 1,48 | 4,2 | 3,54 | 15,82 | 2,02 |
| ***NMHU_07*** | 1,28 | 3,6 | 3,04 | 14,28 | 1,88 |
| ***NMHU_08*** | 1,02 | 2,86 | 2,6 | 13,12 | 1,78 |
| ***NMHU_09*** | 1,04 | 3,14 | 2,76 | 13,54 | 1,74 |
| ***NMHU_10*** | 1,06 | 3,7 | 2,88 | 13,74 | 1,88 |
| ***NMHU_11*** | 1,42 | 4,02 | 3,28 | 15,56 | 2,1 |
| ***NMHU_12*** | 1,22 | 4,22 | 3,16 | 15,6 | 2,18 |
| ***NMHU_13*** | 1,38 | 4,02 | 3,14 | 14,86 | 1,86 |
| ***NMHU_14*** | 1,38 | 3,94 | 2,66 | 13,54 | 1,94 |
| ***NMHU_15*** | 1,22 | 3,6 | 3,24 | 14,08 | 1,58 |

## Fémur

Todos los fémures (NMF_01-NMF_15) corresponden a ejemplares adultos y son de lateralidad derecha. Presentan un buen estado de conservación, con la presencia de los caracteres morfológicos por región hueso aptos para el análisis osteométrico. En la epífisis proximal se conservan: cabeza (**Cab**), trocánter mayor (**TrMy**), trocánter menor (**TrMe**) y la cresta intertrocantérica (**CIntr**). En su región distal preservan: los cóndilos externos (**ConExt**) e internos (**ConInt**).

**TABLA VII.** Parámetros osteométricos conservados para los fémures de *N. micrus*

| ***PARÁM. OSTEOM.*** | **AnDis** | **AnProx** | **DMyC** | **LInter** | **LT** | **PfDf** |
|---|---|---|---|---|---|---|
| ***NMF_01*** | 3,42 | 4,4 | 2,42 | 4,62 | 18,76 | 1,72 |
| ***NMF_02*** | 3,14 | 3,8 | 2,04 | 3,72 | 16,3 | 1,48 |
| ***NMF_03*** | 3,34 | 3,74 | 2,22 | 4 | 17,78 | 1,58 |
| ***NMF_04*** | 3,58 | 3,84 | 2,18 | 4,22 | 18,94 | 1,72 |
| ***NMF_05*** | 3,3 | 3,52 | 2,3 | 3,82 | 16,02 | 1,56 |
| ***NMF_06*** | 3,34 | 4 | 2,18 | 4,12 | 17,2 | 1,38 |
| ***NMF_07*** | 3,64 | 3,66 | 2,02 | 3,9 | 16,42 | 1,74 |
| ***NMF_08*** | 3,46 | 4 | 2,2 | 4,28 | 17 | 1,58 |
| ***NMF_09*** | 3,28 | 3,84 | 2,16 | 4,22 | 16,4 | 1,48 |
| ***NMF_10*** | 3,3 | 3,8 | 2,06 | 4,02 | 16,18 | 1,48 |
| ***NMF_11*** | 3,3 | 3,78 | 2,02 | 3,94 | 15,92 | 1,38 |
| ***NMF_12*** | 3,4 | 3,74 | 2,18 | 3,64 | 16,86 | 1,4 |
| ***NMF_13*** | 3,64 | 3,98 | 2,38 | 4,72 | 19,34 | 1,62 |
| ***NMF_14*** | 3,14 | 3,64 | 2,06 | 3,88 | 16,18 | 1,64 |
| ***NMF_15*** | 3,44 | 3,94 | 2,36 | 4,22 | 17,76 | 1,52 |

GZ#2: **Piezas óseas de *Boromys torrei***

**Hemimandíbulas**

**TABLA VIII.** Caracteres morfológicos conservados (celdas sombreadas) para las hemimandíbulas de *B. torrei.* T=Completos; P=Parcial.

| ***CARAC. MORF.*** | ***BTHM_01*** | ***BTHM_02*** | ***BTHM_03*** | ***BTHM_04*** | ***BTHM_05*** | ***BTHM_06*** |
|---|---|---|---|---|---|---|
| **CrMas** | T | T | T | T | T | T |
| **ForMb** | T | T | T | T | T | T |
| **ForRet** | T | T | T | T | T | T |
| **PrAn** | T | T | | T | P | T |
| **PrMen** | T | T | | T | | |
| **Con** | T | T | | T | T | P |
| **PrCon** | T | T | | | T | T |
| **PrCor** | T | T | | T | T | T |
| **Dst** | T | T | T | T | T | T |
| **EsSg** | T | T | | | T | T |
| **PrPstCon** | T | T | | | T | |
| **Sf** | | | | | | |
| **Inc** | | | | | | |
| **P** | | | | | | |
| **$M^1$** | | | | | | |
| **$M^2$** | | | | | | |
| **$M^3$** | | | | | | |

**TABLA IX.** Parámetros osteométricos conservados para las hemimandíbulas de *B. torrei.*

| *PARÁM. OSTEOM.* | *BTHM_01* | *BTHM_02* | *BTHM_03* | *BTHM_04* | *BTHM_05* | *BTHM_06* |
|---|---|---|---|---|---|---|
| **AlCon** | 10,50 | 9,38 | - | - | 8,64 | - |
| **AlCor** | 9,96 | - | - | - | 7,50 | - |
| **AlCorp** | 5,88 | - | 6,22 | 6,08 | 4,88 | 4,04 |
| **AnCorInc** | 1,20 | 0,28 | 1,08 | 0,90 | - | - |
| **AnMyRmAs** | 6,98 | 6,10 | - | - | 5,92 | 7,18 |
| **AnMeRmAs** | 3,64 | 3,06 | - | - | 2,88 | |
| **LAlvMol** | - | - | - | - | 7,02 | - |
| **LAns** | - | - | - | 23,34 | - | - |
| **LCos** | 23,54 | 20,98 | - | - | 19,36 | - |
| **LCorSMol** | 8,76 | 7,40 | 7,96 | 8,18 | - | 7,70 |
| **LDia** | 5,24 | 3,94 | 4,06 | 4,54 | 3,30 | 4,52 |
| **LSf** | 8,14 | 7,20 | - | 7,42 | 6,40 | - |

## Húmeros

Todos los húmeros (BTHU_01 – BTHU_05) corresponden a ejemplares adultos. Las piezas BTHU_01 – BTHU_04 son de lateralidad izquierda, mientras que la pieza BTHU_05 es de lateralidad derecha. Los caracteres morfológicos conservados por región del hueso permiten un adecuado análisis osteométrico. En la región de la epífisis proximal se conservan: cabeza (**Cab**) y cresta deltoidea (**CrDel**). En la región de la epífisis distal se conservan: fosa coronoidea (**FsCor**) y fosa oleocraneana (**FsOlcr**).

**TABLA X.** Parámetros osteométricos conservados para los húmeros de *B. torrei*

| *PARÁM. OSTEOM.* | *BTHU_01* | *BTHU_02* | *BTHU_03* | *BTHU_04* | *BTHU_05* |
|---|---|---|---|---|---|
| **AnDf** | 1,92 | 1,84 | 1,82 | 1,74 | 1,92 |
| **AnDis** | 4,48 | 4,46 | 5,16 | 4,60 | 5,12 |
| **AnProx** | 4,22 | 4,82 | 4,66 | 4,40 | 4,52 |
| **LDel** | 8,58 | 8,64 | 8,78 | 8,44 | 9,56 |
| **LT** | 19,52 | 19,98 | 20,36 | 19,02 | 21,08 |
| **PfDel** | 3,94 | 3,92 | 4,26 | 3,78 | 4,56 |

## Fémur

Debido al alto grado de fragmentación del material óseo, en este caso solo fue posible el análisis de los caracteres morfológicos y parámetros osteométricos en una sola pieza (BTF_01). Esta pieza presenta lateralidad derecha y conserva la epífisis proximal y la región medial de la diáfisis. En la región de la epífisis proximal conserva: cabeza (**Cab**), trocánter menor (**TrMe**), y el trocánter mayor (**TrMy**). Las medidas realizadas en esta pieza fueron: Anchura proximal (**AnProx**= 6,52 mm), Diámetro mayor de la cabeza (**DMyCab**= 3,02 mm) y Longitud intertrocantérica (**LIntr**= 5,88 mm).

## GZ#2: **Piezas óseas de *Boromys offella***

## Hemimandíbulas

BOHM_01- Esta pieza presenta lateralidad izquierda, y conserva la cresta masetérica (**CrMas**), el premolar (**P**), el primer molar (**M$^1$**), el foramen mandibular (**ForMb**), la fosa retromolar (**FsRet**), el proceso condiloides (**PrCon**), parte del cóndilo (**Con**), parte del proceso coronoides (**PrCor**), el proceso mental (**PrMen**), la diastema (**Dst**), la escotadura sigmoidea (**EsSg**). Las medidas realizadas en esta pieza fueron: Anchura menor de la rama ascendente (**AnMeRmAs**= 5,28 mm), Longitud coronaria de la serie molariforme (**LCorSMol**= 10,10 mm), Longitud de la diastema (**LDia**= 7,32 mm) y Longitud de la sínfisis (**LSf**= 11,40 mm).

## Húmeros

En el caso particular de los húmeros de esta especie solo se consideró el estudio de una sola pieza (BFHU_01) debido a su buen estado de conservación, independientemente de su condición de subadulto. La pieza presenta una lateralidad derecha, y conserva los caracteres morfológicos y los parámetros osteométricos adecuadamente, con excepción de su extremo más proximal (caput). En su región medio-proximal está presente la cresta deltoidea (**CrDel**). En la región de la epífisis distal conserva la fosa coronoidea (**FsCor**) y la fosa olecraneana (**FsOlcr**). Las mediciones realizadas a esta pieza fueron: Anchura de la diáfisis (**AnDf**= 2,64 mm), Anchura distal (**AnDis** = 7,66 mm), Profundidad deltoidea (**PfDel**= 6,34 mm), aún sin la total fusión del capú esta pieza presenta una longitud total (**LT**) de 29,92 mm.

**Fémur**

**TABLA XI.** Caracteres morfológicos conservados (celdas sombreadas) para los fémures de *B. offella.* T=Completos; P=Parcial.

| *CARAC. MORF.* | *BOF_01* | *BOF_02* | *BOF_03* | *BOF_04* | *BOF_05* |
|---|---|---|---|---|---|
| **ConExt** | T | T | | | |
| **TrMy** | T | T | T | P | T |
| **TrMn** | T | T | T | T | T |
| **Cab** | T | T | T | T | T |
| **Df** | T | T | T | | |
| **FsLigR** | T | T | T | T | T |

**TABLA XII.** Parámetros osteométricos conservados para los fémures de *B. offella.*

| *PARÁM. OSTEOM.* | *BOF_01* | *BOF_02* | *BOF_03* | *BOF_04* | *BOF_05* |
|---|---|---|---|---|---|
| **AnDf** | 2,74 | 3,22 | 3,12 | - | - |
| **AnDis** | 5,36 | 5,48 | - | - | - |
| **AnProx** | 6,98 | 6,78 | 7,76 | - | 7,52 |
| **DMyCab** | 3,14 | 2,82 | 3,20 | 3,06 | 3,22 |
| **LIntr** | - | 7,02 | 6,04 | - | 6,44 |
| **LT** | 28,96 | 31,72 | - | - | - |
| **PfDf** | 2,08 | 2,32 | 2,30 | - | - |

GZ#2: **Piezas óseas de *Geocapromys columbianus.***

**Hemimandíbulas**

**TABLA XIII.** Caracteres morfológicos conservados (celdas sombreadas) para las hemimandíbulas de *G. colombianus.* T=Completos; P=Parcial.

| *CARAC. MORF.* | *GCHM_01* | *GCHM_02* | *GCHM_03* | *GCHM_04* | *GCHM_05* |
|---|---|---|---|---|---|
| **CrMas** | P | P | T | P | P |
| **ForMb** | | | | | |
| **FsRet** | | | | | |
| **PrAn** | | | | | |
| **PrMen** | T | T | T | T | T |
| **Con** | | | | | |
| **PrCon** | | | | | |
| **PrCor** | P | | | | |
| **Dst** | | | | | |
| **EsSg** | | | | | |
| **PrPstCon** | | | | | |
| **Sf** | | | | | |
| **Inc** | | | | | |
| **P** | | | | | |
| **$M^1$** | | | | | |
| **$M^2$** | | | | | |
| **$M^3$** | | | | | |

GZ#2: **Piezas óseas de *Mesocapromys* sp.**

**Hemimandíbula**

Este género presenta una sola pieza ósea (MCHM_01) de lateralidad izquierda. Su estado de conservación no permitió determinar la especie en particular. Solamente conserva la cresta masetérica (**CrMas**) en estado de fragmentación, conserva además el premolar (**P**) y el primer molar (**$M^1$**). Las mediciones realizadas a esta pieza fueron: Altura corporal (**AlCorp**= 6,20 mm) y Longitud de la Sínfisis (**LSf**= 8,40 mm).

GZ#3: **Piezas óseas de *Brachyphylla nana***

**Hemimandíbulas**

Esta especie está representada por dos piezas óseas (BNHM_01, BNHM_02) de lateralidad izquierda. Ambas piezas se encuentran fragmentadas en su porción anterior, con ausencia de la región de la sínfisis mandibular (**SfMb**), foramen mentoniano (**ForMen**), e incisivos (**Inc**). La primera conserva el proceso coronoides (**PrCor**), el cóndilo mandibular (**ConMb**) y el proceso angular (**PrAn**). En la segunda pieza el proceso coronoides (**PrCor**) aparece fragmentado, y se conservan el cóndilo mandibular (**ConMb**), el proceso angular (**PrAn**), el tercer premolar (**$P^3$**) y el tercer molar (**$M^3$**).

GZ#3: **Piezas óseas de *Erophylla sezekorni***

Especie representada por dos piezas óseas (ESHM_01, ESHM_02), ambas de lateralidad izquierda. Estas piezas se encuentran fragmentadas en su porción anterior, con ausencia de la región de la sínfisis mandibular (**SfMb**), foramen mentoniano (**ForMen**), e incisivos (**Inc**). La primera conserva el proceso coronoides (**PrCor**), el cóndilo mandibular (**ConMb**) y el proceso angular (**PrAn**). La segunda pieza ósea presenta el proceso coronoides (**PrCor**) y el proceso angular (**PrAn**) fragmentados; solamente conserva el cóndilo mandibular (**ConMb**) y el tercer premolar (**$P^3$**).

GZ#3: **Piezas óseas de *Phyllonycteris poeyi***

Especie representada por dos piezas óseas (PPoHM_01, PPoHM_02), ambas de lateralidad izquierda y fragmentadas en su porción anterior con ausencia de la región de la sínfisis mandibular (**SfMb**), e incisivos (**Inc**). También en ambas piezas se conservan los caracteres osteológicos: foramen mentoniano (**ForMen**), proceso coronoides (**PrCor**), cóndilo mandibular (**ConMb**) y proceso angular (**PrAn**).

GZ#3: **Piezas óseas de *Macrotus waterhousei***

**TABLA XIV.** Caracteres morfológicos conservados (celdas sombreadas) para las hemimandíbulas de *M. waterhousei.* T=Completos; P=Parcial.

| ***CARAC. MORF.*** | **ForMen** | **SfMb** | **PrCor** | **ConMb** | **PrAn** | **Inc** | **Can** | **P¹** | **P²** | **M¹** | **M²** | **M³** |
|---|---|---|---|---|---|---|---|---|---|---|---|---|
| ***MWHM_01*** | T | T | T | T | T | | | | | | | |
| ***MWHM_02*** | T | T | T | T | T | | | | | | | |
| ***MWHM_03*** | T | T | T | T | T | | | | | | | |
| ***MWHM_04*** | T | T | T | T | T | | | | | | | |
| ***MWHM_05*** | T | T | T | T | | | | | | | | |
| ***MWHM_06*** | T | T | T | T | T | | | | | | | |
| ***MWHM_07*** | T | T | T | T | T | | | | | | | |
| ***MWHM_08*** | T | T | T | T | T | | | | | | | |
| ***MWHM_09*** | T | T | T | T | T | | | | | | | |
| ***MWHM_10*** | T | T | T | T | T | | | | | | | |
| ***MWHM_11*** | T | T | T | T | | | | | | | | |

Las piezas MWHM_01 – MWHM_05 presentan una lateralidad izquierda, el resto de las piezas (MWHM_06 – MWHM_11) son de lateralidad derecha.

GZ#3: **Piezas óseas de *Monophyllus redmani***

Esta especie está representada por una sola pieza ósea (MRHM_01), de lateralidad izquierda, la cual conserva el foramen mentoniano (**ForMen**), el proceso coronoides (**PrCor**), el cóndilo mandibular (**ConMb**) y el proceso angular (**PrAn**).

GZ#3: **Piezas óseas de *Eptesicus fuscus***

Especie representada por dos piezas óseas (EFHM_01, EFHM_02), ambas de lateralidad izquierda, conservando el foramen mentoniano (**ForMen**), el cóndilo mandibular (**ConMb**) y el proceso angular (**PrAn**). Particularmente la primera pieza presenta el proceso coronoides (**PrCor**) fragmentado. Las medición realizada a esta pieza fue: la longitud mentoniana (**LM**= 11,54 mm). La segunda presenta el proceso coronoides (**PrCor**) completo, además del primer ($\mathbf{M^1}$) y segundo molar ($\mathbf{M^2}$). Las medición realizada a esta pieza fue: la longitud mentoniana (**LM**= 10,02 mm).

GZ#3: **Piezas óseas de *Pteronotus parnelli***

Especie representada por dos piezas óseas (PPHM_01, PPHM_02), de lateralidad derecha. La primera pieza ósea conserva el foramen mentoniano (**ForMen**), el proceso coronoides (**PrCor**), el cóndilo mandibular (**ConMb**), el proceso angular (**PrAn**) y el segundo molar ($\mathbf{M^2}$). La segunda pieza ósea se encuentra fragmentada en su porción anterior, con ausencia de la región de la sínfisis mandibular (**SfMb**), foramen mentoniano (**ForMen**), e incisivos (**Inc**), conservando solamente el proceso coronoides (**PrCor**), el cóndilo mandibular (**ConMb**) y el proceso angular (**PrAn**).

GZ#3: **Piezas óseas de *Tadarida brasiliensis***

Especie representada por dos piezas óseas (TBHM_01, TBHM_02), ambas de lateralidad derecha. La primera pieza ósea conserva el foramen mentoniano (**ForMen**), el proceso coronoides (**PrCor**), el cóndilo mandibular (**ConMb**), y el proceso angular (**PrAn**); mientras que la segunda pieza ósea se encuentra fragmentada en su porción anterior, con ausencia de la región de la sínfisis mandibular (**SfMb**), foramen mentoniano (**ForMen**), e incisivos (**Inc**). No obstante esta pieza conserva el proceso coronoides (**PrCor**), el cóndilo mandibular (**ConMb**), el proceso angular (**PrAn**) y el segundo ($\mathbf{M^2}$) y tercer molar ($\mathbf{M^3}$).

GZ#3: **Piezas óseas de *Phyllops silvai***

Esta especie está representada por una sola pieza ósea (PSHM_01), de lateralidad izquierda. La misma conserva el foramen mentoniano (**ForMen**), el proceso coronoides (**PrCor**), el cóndilo mandibular (**ConMb**), el proceso angular (**PrAn**) y el segundo premolar (**$M^2$**).

## CAPÍTULO 4. DISCUSIÓN

### 4.1. Consideraciones taxonómicas sobre el depósito paleontológico.

La fauna de mamíferos cubanos, vivientes y extinguidos, ha sido una de las más estudiadas en los últimos tiempos, en la cual los trabajos taxonómicos se han actualizado considerablemente (Silva *et al.* [2008]; Borroto-Páez y Mancina, 2011). Esto constituye una ventaja desde hace muchísimos años con respecto a otros grupos de vertebrados, tales como los anfibios y los reptiles fósiles (paleoherpetofauna), que siguen siendo los menos conocidos a raíz de la carencia de especialistas encargados en generar y procesar materiales de referencia y, a su vez, estudios asociados a los mismos.

Por otra parte, las investigaciones acerca de la ornitofauna fósil también ocupan un lugar importante entre los principales resultados de la paleontología cubana (Suárez, 2005), entre las que se destaca el estudio de residuarios producto de la acción de especies depredadoras, tanto de hábitos diurnos como nocturnos (Jiménez-Vázquez *et al.,* 2005).

Son precisamente estas investigaciones las que han tenido que acudir a la interacción multidisciplinaria de especialistas en función de la identificación de los taxones presentes en los mismos. La composición taxonómica de cualquiera de estos residuarios puede abarcar diversos grupos zoológicos dentro de los vertebrados, tanto extintos como vivientes (Arredondo y Chirino, 2002; Hernández-Muñoz y Mancina, 2011; López-Ricardo y Borroto-Páez, 2012). El análisis de la composición de los depósitos originados por la acción depredadora de lechuzas juega un papel esencial en el conocimiento ecológico de especies actuales y pasadas, aportando elementos acerca de su distribución, densidad poblacional, y en algunos casos, puede ser un indicador de extinción de algunas especies que desaparecen de sus registros alimentarios, y de los estratos a los que éstos se encuentran asociados.

El estudio de la composición taxonómica del depósito "El Abrón" refuerza lo anteriormente expuesto, y además, no constituye simplemente un inventario taxonómico más de la fauna de vertebrados cubanos, sino que se trata de un residuario producido a lo largo del tiempo por dos especies de lechuzas (*Tyto noeli* y *Tyto alba*) esencialmente, donde una de ellas se encuentra extinta (*T. noeli*), y se considera responsable de las acumulaciones más antiguas en la estratigrafía del depósito.

La importancia de este aspecto radica además en que dos niveles por encima del Nivel IX (objeto de estudio de este trabajo, ver Fig. 1) se hayan fechado restos óseos de esta especie, que ubica su acción trófica, así como la fauna asociada al estrato en que apareció en unos 17 406 ±161 años AP (Suárez y Díaz-Franco, 2003), lo que sugiere una antigüedad mayor para las especies identificadas y aporta nuevos elementos para las reconstrucciones paleoecológicas del Cuaternario de Cuba.

Un ejemplo particularmente importante, además de los restos óseos fechados de *Tyto noeli*, lo constituye el hecho de la identificación de una nueva especie de murciélago para Cuba (*Phyllops silvai*), procedente del estrato de mayor grosor (Nivel VII, ver Fig. 1), en el trabajo de Suárez y Díaz-Franco (2003), constituyendo este depósito su localidad tipo. En el presente trabajo se expone como resultado la identificación de una hemimandíbula izquierda (PSHM_01) de esta especie de murciélago, extraída de algún punto de la estratigrafía del Nivel IX, la cual representa el registro más antiguo para la misma, si asumimos una antigüedad mayor de todos los restos del Nivel IX respecto al Nivel VII.

Lamentablemente en los trabajos previos realizados en este depósito no se contempló la ubicación exacta en el espesor de la capa estratigráfica de los restos extraídos, dígase tanto para el murciélago extinto *Phylllops silvai* como para el ave extinta *Tyto noeli*. Puede esto estar relacionado con la altísima densidad de restos acumulados, considerados en millones (Suárez y Díaz-Franco, 2011), y sus características de deposición, lo cual requeriría de técnicas de excavación extremadamente minuciosas, y un tiempo muy prolongado de trabajo de campo, aspectos estos que desconocemos si estaban al alcance de los investigadores que llevaron a cabo las labores paleontológicas entre los años 2000-2001.

Otro elemento taxonómico a considerar es la presencia de otras especies dentro del orden Chiroptera, no reportadas para este municipio en estudios anteriores (Tabla I). Sin embargo, conocemos que la amplia distribución de estas especies a lo largo del archipiélago cubano (Silva, 1979) justifica ampliamente la presencia en el depósito de cualquiera de las reportadas en este estudio. Si observamos la Fig. 7, podemos apreciar que, a pesar de que los puntos de distribución no se encuentran exactamente en el municipio Los Palacios, es muy evidente su significativa distribución en la provincia, e incluso en áreas bien cercanas al depósito.

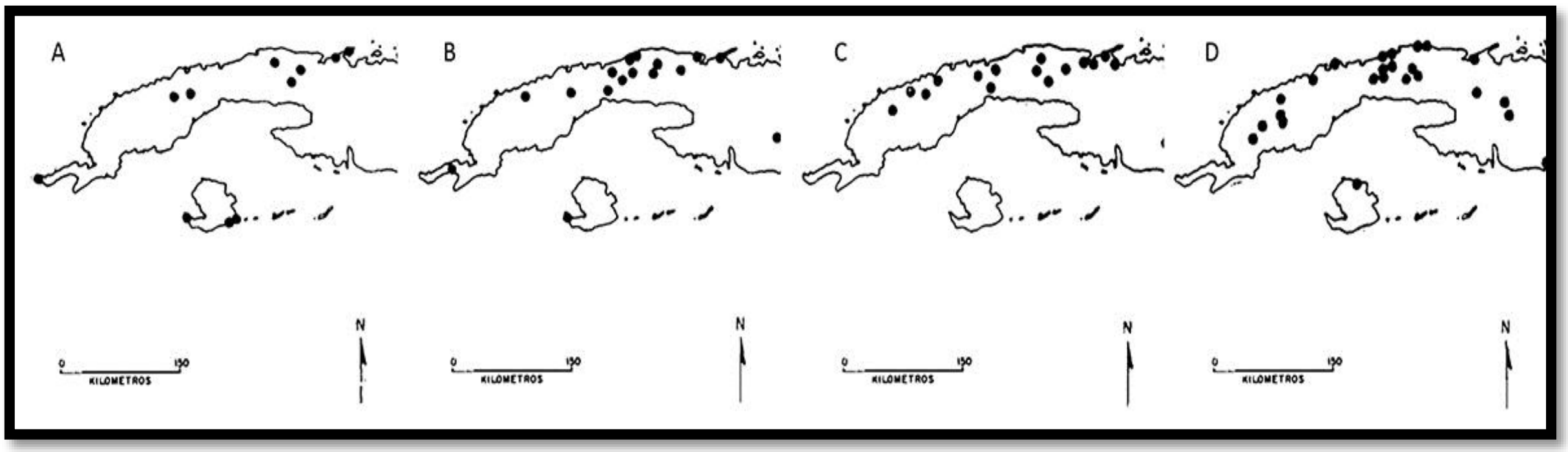

**FIGURA 7**. Distribución de las especies de murciélagos (**A**) *Erophylla sezekorni,* (**B**) *Monophyllus redmani,* (**C**) *Pteronotus parnelli, y* (**D**) *Tadarida brasiliensis* en la región más occidental de Cuba. Tomado y modificado de Silva (1979).

Asimismo, el género *Mesocapromys* dentro del orden Rodentia constituye un punto de análisis importante en el presente trabajo. Para este caso, la pieza ósea analizada (MCHM_01) no ofrece toda la información necesaria para la identificación, hasta el momento, de la especie en particular, según el criterio propuesto en Silva *et al.* [2008]. La determinación del género estuvo basada en el criterio de Arredondo (2015*, com. pers*.) además de las características del diseño oclusal del $\mathbf{M^1}$, propuesta en Silva *et al.* [2008].

No obstante, si tenemos en cuenta algunos datos acerca de las especies de este género en el trabajo antes citado se puede arribar a determinadas consideraciones tales como: (1) las especies conocidas hasta ahora se resumen en cinco taxones distribuidos diferentemente por el archipiélago cubano (véase Silva *et al.* [2008:153-156]; (2) la especie *Mesocapromys nanus* es la única actualmente reportada para la provincia de Pinar del Río; por lo que (3) pudiese sugerirse de que se trate de la especie mencionada.

No obstante, la pieza ósea ofrece medidas enteramente por debajo de todas las especies del género, incluyendo *M. nanus*. El estado de conservación de la pieza no permite delimitar con precisión si se trata de un individuo subadulto, aunque es totalmente aceptable esta posibilidad. El otro aspecto es que para *M. nanus* en Silva *et al.* [2008] se muestran valores osteométricos para la especie que habita actualmente y la especie en estado fósil. Los valores de la pieza (MCHM_01) se mantienen por debajo de ambas, aunque son más parecidos a los determinados para la especie fósil (véase Silva *et al.* [2008:194]).

También dentro del orden Rodentia, el último de los taxones a considerar en este análisis es la especie *Geocapromys columbianus*, la cual también presenta una amplia distribución por todo el registro fósil cubano, y donde la provincia de Pinar del Río tiene una considerable representación. En principio, en el trabajo de Silva *et al.* [2008] no se hace referencia en el catálogo de localidades acerca de la presencia de esta especie en el área donde se ubica el depósito. Esto trae consigo de que pueda considerarse la especie en este estudio como un nuevo reporte para esta localidad, aunque cabe señalar que en el reciente trabajo de Suárez y Olson (2015) los autores hacen mención a este yacimiento durante su análisis de la distribución y actualización de especies del género *Tyto* (Clase Aves), destacando que juveniles de especies de micromamíferos hallados en “El Abrón” tales como *G. columbianus* y *B. offella* constituyen la principal dieta de estas aves depredadoras.

En relación con lo expuesto anteriormente también es importante resaltar que la abundancia de material óseo perteneciente a individuos juveniles de mamíferos es un aspecto importante en el análisis de los hábitos alimentarios de la lechuza, los cuáles han demostrado que son muy importantes en el conocimiento sobre la fauna actual, pero sobre todo, la del pasado, dentro de  la cual muchas especies extintas se asocian a este tipo de actividad de depredación.

Finalmente, como el estudio de este depósito fosilífero aún se encuentra en desarrollo, algunos niveles por debajo del nivel fechado carecen de suficiente investigación. Solamente algunas observaciones al respecto se aprecian en informes científico-técnicos (ICT) del Departamento de Paleogeografía y Paleobiología del Museo Nacional de Historia Natural, de La Habana, Cuba (MNHNHCu), donde se destaca la aparición de taxones introducidos en los estratos más recientes, asociados a la acción de la lechuza *Tyto alba*; observaciones éstas asociadas a las labores de extracción, y limpieza de los materiales más que a un análisis paleontológico exhaustivo. No obstante, estos señalamientos están en total correspondencia con los estudios actuales acerca de la composición taxonómica de residuarios de lechuzas para nuestro archipiélago (Jiménez-Vázquez *et al.,* 2005; Hernández-Muñoz y Mancina, 2011; López-Ricardo y Borroto-Páez, 2012).

**4.2. Consideraciones tafonómicas del depósito.**

Las características del depósito resumidas en los trabajos previos realizados para “El Abrón” (Suárez y Díaz-Franco, 2003; 2011) apuntan sin duda alguna a una clasificación del origen del yacimiento Tipo A, según los criterios de Silva (1974), y Woloszyn y Silva (1977). Este tipo de formación originados por la depredación de la lechuza *(Tyto)*, constituyen los restos de su actividad trófica, donde procesos de acumulación gradual de restos y transporte por acción hidráulica pueden intervenir en la acumulación final de los materiales producidos biogénicamente por la entidad paleobiológica productora.

El primer aspecto a considerar es que uno de los elementos que señalan Woloszyn y Silva (1977) es que este tipo de depósito generalmente tiende a no conservarse, debido a la cercanía de las acumulaciones de las estrígidas a las entradas de las cuevas o solapas cársicas, donde los efectos del intemperismo actúan destructivamente sobre la acumulación.

Sin embargo, el fechado mencionado con anterioridad en el antepenúltimo de sus niveles, a 0,90 cm del sustrato final del depósito, agrega sin duda un punto fuerte al análisis de este tipo de yacimiento, redefiniéndolo en el sentido de sus posibilidades de conservación de los materiales generados por estrígidas. No cabe duda de que la generalidad se corresponde con lo propuesto por Woloszyn y Silva (1977), pero tanto los estudios previos en este depósito (Suárez y Díaz-Franco, 2003) así como este trabajo, se aportan elementos suficientes para considerar caracteres mixtos en el mismo (Tipo A-B, *en* Woloszyn y Silva, 1977). En ninguno de los trabajos mencionados que tratan del depósito objeto de estudio se habla de alteraciones antropogénicas (con excepción de las recientes) en los estratos acumulados, por lo que se considera origen natural para los procesos de deposición.

La importancia de conocer los procesos que le dieron origen a los depósitos así como los mecanismos que incidieron en los estados de conservación de los materiales paleontológicos es que permite evaluar las condiciones de los restos óseos, conocer qué procesos paleoambientales pudieron ser responsables de tales transformaciones o modificaciones, además de que evita la sobredimensión taxonómica de especies o equivocaciones en la identificación, si se realiza un análisis adecuado en este sentido, denominado *tafonómico*.

Para nuestro caso particular, un ejemplo a considerar fue la valoración de las piezas óseas de *N. micrus* y las especies de murciélagos, las cuales presentan un parecido extraordinario, sobre todo en sus piezas hemimandibulares. Como puede corroborarse en la Tabla I, así como en el resto de las descripciones realizadas para las piezas óseas de estas especies, la ausencia de determinados caracteres e incluso, regiones completas de los huesos es una característica bastante común.

Estos niveles de fragmentación se corresponden con la manera en que se han acumulado los restos. La producción del tipo *biogénica* (Fernández-López, 2000a) se justifica a partir de que se conoce la entidad paleobiológica productora (lechuzas) de casi la totalidad de los restos acumulados en cada estrato. Partimos de que durante el proceso de ingestión y regurgitación de que llevan a cabo las estrígidas suceden los primeros eventos de fragmentación, donde además, la acción gástrica de sus estómagos [glandular o proventrículo, y muscular o ventrículo (molleja)] añade pequeños efectos de disolución sobre los tejidos, que contribuye al deterioro de algunas piezas óseas posterior al proceso de regurgitación, donde muchas de las egagrópilas tienden a desintegrarse. Algunas consideraciones acerca de todo el proceso antes mencionado puede encontrarse en el trabajo de López-Ricardo y Borroto-Páez (2012).

Tal como señalan Suárez y Olson (2015) al constituir los individuos juveniles de la microfauna las principales presas para estas estrígidas no podemos obviar los factores intrínsecos como densidad ósea y niveles de osificación, que añaden vulnerabilidad a las piezas esqueléticas, contribuyendo a su fragmentación, y en el peor de los casos, su entera desaparición.

La asociación de fósiles que se encuentra en el depósito "El Abrón" se reconoce como *orictocenosis*, definida como "*conjunto de fósiles que están, o han sido encontrados juntos"* (Fernández-López, 2000a:34). Atendiendo a las características de los materiales óseos y su estado actual de conservación, es posible que los materiales acumulados en este depósito respondan también a procesos conocidos como *resedimentación* y *reelaboración tafonómica*, los cuales son consecuencia de alteraciones tanto ambientales como estratigráficas de los restos, en ocasiones, antes del proceso final de la acumulación. Sin embargo en este depósito, naturalmente, los restos acumulados no se corresponden con la manera en que se presentan los residuos actuales de egagrópilas, donde gran parte de los componentes óseos de las especies aparecen aún relacionados anatómicamente. Esto se debe a la antigüedad del sitio y a la desintegración paulatina de las egagrópilas, quedando los restos totalmente vulnerables a la desarticulación y la dispersión de sus elementos, perdiéndose toda asociación anatómica entre los restos acumulados. Para este tipo de fenómeno tafonómico (ausencia total de esqueletos completamente o parcialmente articulados, huesos dispersos, aislados, etc.) Holz y Barberena (1994) proponen clasificar yacimientos con estas características como **Clase III** con las subclases **a** y **b**.

Independientemente del origen biogénico de los restos acumulados, muchas piezas óseas bajo la acción de los mecanismos de alteración tafonómica responsables de procesos como la fragmentación, generan a partir de una sola pieza ósea varios elementos independientes. Este tipo de "multiplicación" de los elementos producidos por entidades paleobiológicas se considera una producción tafogénica, la cual, acompañada de procesos de desarticulación y dispersión, contribuye con la densidad de materiales y con la alta fragmentación de los restos que le dieron origen.

Otro factor tafonómico a considerar es la compresión por carga litostática, el cual, atendiendo a la estratigrafía presentada en la Fig. 1 del presente trabajo, se considera que pudo haber incidido en los niveles de fragmentación y deterioro de muchas piezas óseas, sobre todo de aquellas contenidas en los niveles estratigráficos más profundos, donde el Nivel IX es el más afectado. Si revisamos el material taxonómico examinado notamos una ausencia casi total de cráneos de cualquiera de las especies analizadas, solo con la excepción de tres piezas muy fragmentados de *N. micrus* (Tabla 1).

## CONCLUSIONES

1. El proceso de identificación de 3 órdenes, 7 familias y 14 especies de mamíferos en el nivel IX del depósito paleontológico "El Abrón" constituye un paso importante en los estudios de microvertebrados del yacimiento, lo que permite futuros estudios e interpretaciones paleoecológicas; en aras de reconstruir segmentos de la historia natural de las especies estudiadas.
2. Teniendo como referencia el fechado de 17 400 años AP obtenido en el Nivel VII, toda la fauna registrada en el depósito, posterior a este nivel, incrementa su antigüedad; además de que el hallazgo de cuatro taxones dentro del orden Chiroptera, y un taxón dentro del orden Rodentia, constituye un aporte al conocimiento acerca de la distribución de estas especies en el archipiélago cubano.
3. El análisis del estado de conservación de los restos óseos sobre la base de criterios tafonómicos facilita la comprensión de los procesos que dieron origen al depósito, y permiten comprender el porqué de los altos niveles de acumulación, dispersión, desarticulación y fragmentación de los materiales paleontológicos. Este conocimiento contribuye a la no sobredimensión de especies y a una correcta relación espaciotemporal de los restos.

## RECOMENDACIONES

1. Extender el estudio paleontológico de microvertebrados realizado en los mamíferos del Nivel IX al resto de los grupos zoológicos presentes en las acumulaciones del depósito (anfibios, reptiles, aves y mamíferos) de los niveles en general, los cuales se encuentran almacenados en el Museo Nacional de Historia Natural, La Habana, Cuba (MNHNCu).
2. Aplicar una metodología de excavación más precisa que permita registrar en cada segmento la ubicación de los taxones más representativos, así como un mejor control de los niveles estratigráficos, que contribuya además a realizar adecuadas valoraciones tafonómicas acerca del origen del depósito y del estado de conservación de los restos óseos.

**BIBLIOGRAFÍA CITADA**